\documentclass[%
 reprint,
superscriptaddress,
 amsmath,amssymb,
 aps,
prx,
]{revtex4-2}

\usepackage{soul,upgreek}
\soulregister{\cite}7
\soulregister{\ref}7
\usepackage{color,xcolor}
\usepackage[T1]{fontenc}
\usepackage{graphicx}
\usepackage{dcolumn}
\usepackage{bm}

\usepackage[final]{changes}

\usepackage[dvipsnames]{xcolor}
\makeatletter
\AddToHook{cmd/added/before}{\def\Changes@AuthorColor{red}}
\AddToHook{cmd/deleted/before}{\def\Changes@AuthorColor{orange}}
\AddToHook{cmd/replaced/before}{\def\Changes@AuthorColor{blue}}
\makeatother

\usepackage[colorlinks,
linkcolor=blue,
anchorcolor=blue,
citecolor=blue,
urlcolor=blue]{hyperref}

\begin{document}
	
	\preprint{APS}
	
	\title{{A Quantum Computer Based on Donor-Cluster Arrays in Silicon}}
	\title{\added{Scalable Spin Qubit Architecture with Donor-Cluster Arrays in Silicon}}
    
	\author{Shihang Zhang}
    \altaffiliation{These authors contributed equally to this work.}
    \affiliation{Southern University of Science and Technology, Shenzhen 518055, China}
	\affiliation{International Quantum Academy, Shenzhen 518048, China}
    \author{Guangchong Hu}
    \altaffiliation{These authors contributed equally to this work.}
    \affiliation{International Quantum Academy, Shenzhen 518048, China}
    \affiliation{Shenzhen Branch, Hefei National Laboratory, Shenzhen 518048, China.}
	\author{Chunhui Zhang}
    \affiliation{Southern University of Science and Technology, Shenzhen 518055, China}
	\affiliation{International Quantum Academy, Shenzhen 518048, China}
    \author{Guanyong Wang}
    \affiliation{International Quantum Academy, Shenzhen 518048, China}
    \author{Tao Xin}
    \affiliation{International Quantum Academy, Shenzhen 518048, China}
	\author{Yu He}
	\email{hey@iqasz.cn}
    \affiliation{International Quantum Academy, Shenzhen 518048, China}
    \affiliation{Shenzhen Branch, Hefei National Laboratory, Shenzhen 518048, China.}
	\author{Peihao Huang}  
	\email{huangpeihao@iqasz.cn}
    \affiliation{International Quantum Academy, Shenzhen 518048, China}

	\date{\today}

	\begin{abstract}
    Spin qubits in silicon donors offer a promising platform for quantum computing due to their long coherence times and semiconductor compatibility. 
    \added{However, scaling donor-based spin qubits in silicon is fundamentally challenged by frequency crowding, crosstalk, and the tight tolerances on donor placement in conventional single-donor architectures.
    To overcome this, we introduce a paradigm based on a two-dimensional array of phosphorus-donor clusters, in which multiple donors share a bound electron. 
    The natural hyperfine distribution within each cluster enables individual addressability of the electron and nuclear spins, while tunable exchange interactions between clusters mediate local all-to-all connectivity. 
    We present a universal control protocol achieving gate fidelities exceeding 99\% for both intra-cluster and inter-cluster multi-qubit operations, with crosstalk effectively suppressed. The architecture natively supports efficient quantum error correction, including bias-tailored codes that exploit the intrinsic noise bias of spin qubits. Furthermore, its modular design is compatible with long-range coupling via electron shuttling for large-scale integration. This donor-cluster array architecture establishes a robust and hardware-efficient pathway towards scalable, fault-tolerant quantum computing in silicon.}
 
	\end{abstract}

	\maketitle
	\section{Introduction}\label{section:intro}

Silicon as a platform for spin-based quantum computing holds significant promise, combining exceptionally long coherence times and inherent compatibility with modern semiconductor manufacturing~\cite{Kane1998,Loss1998,tyryshkin2012,Zwanenburg2013,Burkard2021, Saeedi2013,Muhonen2014,HuangJY2024,steinacker2025,mcintyre2026theory}. These attributes have driven remarkable experimental advances, including the demonstration of single-qubit and two-qubit gate fidelities exceeding 99.99\% and 99.9\%~\cite{wuYh2025,edlbauer2025}. Scalability prospects are further strengthened by recent achievements such as CMOS-compatible control electronics, high-fidelity gate at temperatures above 1 Kelvin, long-distance spin shuttling via moving quantum dots, cryo-CMOS multiplexers, and the realization of multi-qubit processors~\cite{yang_operation_2020,Takeda2021,ruffino2022,Philips2022,WangCA2024,HuangJY2024,edlbauer2025,wuYh2025,mcintyre2026theory,ademi2025distributing,dijkema2026simultaneous,undseth2026weight,tosato_crossbar_2026}. Together, these developments make it attractive to realize quantum computing in silicon.

A promising pathway for silicon quantum computing is offered by donor spin qubits, such as the $^{31}$P donor in silicon~\cite{Kane1998, Zwanenburg2013}. Donors provide two well-defined qubits—the electron and the nuclear spin ($I = 1/2$) with near-zero leakage, zero loss due to the stable dopant, and exceptionally long coherence times, particularly for the nuclear spin~\cite{Muhonen2014}. 
In addition, donor spins benefit from atomic-precision placement via STM lithography~\cite{Fuechsle2012,Zwanenburg2013} and permit high-fidelity control~\cite{Muhonen2015, Madzik2022,Thorvaldson2024}. The electron spin resonance (ESR) enables multi-qubit gates among nuclear spins coupled to a shared electron~\cite{Madzik2022, Thorvaldson2024, edlbauer2025}, while nuclear magnetic resonance (NMR) allows the single-qubit gate of nuclear spins. Moreover, the exchange coupling between electrons of neighboring donors can reach the megahertz regime, and mediate the entanglement of nuclear spins~\cite{edlbauer2025,stemp2025}. 


\added{Despite these achievements, scaling donor-based quantum processors within the prevailing single-donor architecture faces fundamental bottlenecks \cite{Hill2015,Tosi2017}. The inherent uniformity of donor atoms leads to frequency crowding and crosstalk, making individual qubit addressability a formidable task. Moreover, the stringent requirement for deterministic, atomic-precision placement poses a major manufacturing challenge for large-scale systems~\cite{wyrick2022}.}
Additional obstacles include \added{readout ionization shocks~\cite{Monir2024} and limited qubit connectivity, which complicates quantum error correction (QEC)}. Although multi-qubit nuclear spin entanglement in donor clusters has been demonstrated recently~\cite{edlbauer2025, Thorvaldson2024, ZhangCH2025,ZhangCH2026}, a complete and scalable architectural framework based on donor clusters for \added{fault-tolerant quantum computing (FTQC), along with a comprehensive performance analysis, remains an open challenge.}

\begin{figure*}[t]
		\begin{center}
			\includegraphics[width=1.9\columnwidth]{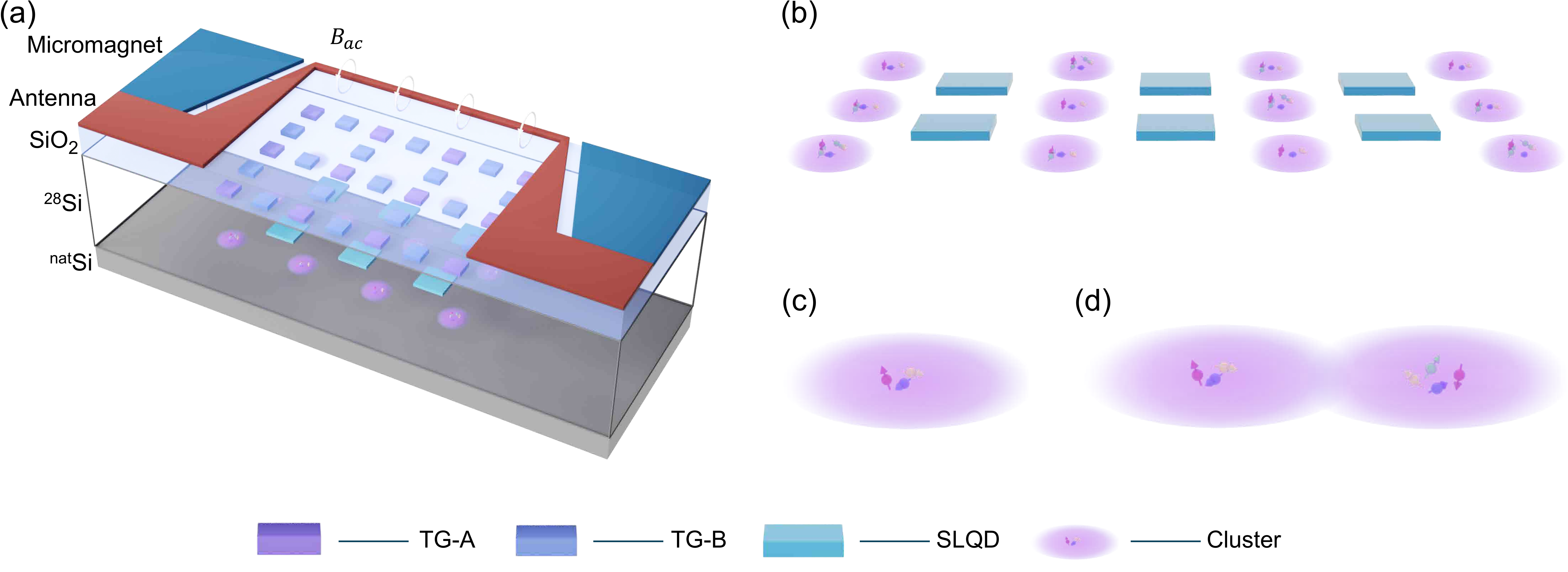}
		\end{center}
		\vspace{-0.50cm}
		\caption{Physical schematic diagram of a quantum processor based on the donor cluster scheme. (a) The donor cluster array is positioned in the $^{28}$Si layer on top of the natural silicon ($^{\rm{nat}}$Si) layer. Each cluster contains a varying number of nuclei (represented by colored arrows) of donors and a shared bound electron (depicted as `cloud’ surrounding the nuclei). The single-lead quantum dots (SLQDs) between clusters, represented as light blue boxes, are utilized for electron loading and readout. Other elements within the cluster layer are controlled by the top gates (TGs) embedded in the upper oxide layer. TG-A is used to adjust the electron energy detuning between clusters, while TG-B is used to modulate the inter-cluster tunneling. At the top of the substrate are the micromagnets that provide the magnetic field gradient and the antennas that generate the alternating magnetic field $B_{ac}$. (b) Zoomed-in view of the cluster layer in $^{28}$Si, where the clusters are arranged in a two-dimensional array. (c) A single cluster comprises multiple nuclei and a shared bound electron. (d) Exchange-coupled two neighboring clusters, where inter-cluster nuclear spin coupling is mediated by the electron-electron exchange interaction.}
		\label{fig:device-2d}
	\end{figure*}


To overcome the challenges, we introduce a cluster-array paradigm for donor-based quantum computing and 
\added{establish a complete pathway from physical realization to FTQC.
(i) We present a scalable donor-cluster array architecture that turns fabrication variability into a functional resource for qubit addressability. (ii) We demonstrate a universal control protocol capable of achieving high gate fidelities exceeding the fault-tolerance (FT) threshold, with the crosstalk error effectively suppressed. (iii) The architecture natively supports efficient, bias-tailored QEC codes for FTQC due to its local all-to-all connectivity and the intrinsic noise bias of spin qubits. 
(iv) We outline a integration roadmap compatible with established techniques like electron shuttling for large-scale implementation. 
By bridging innovative design with rigorous analysis, this work establishes a robust and scalable pathway toward FTQC in silicon.
}

	\section{Quantum computing architecture based on donor-cluster array}\label{sec: scalable scheme}
	
    The architecture of our proposed cluster-based spin qubit scheme is shown in Fig.~\ref{fig:device-2d}(a). Clusters are arranged in a 2D array. Each donor cluster contains multiple P donors and one shared electron in $^{28}$Si, where the nuclear (or electron) spin acts as the data (or ancilla) qubit. As shown in Fig.~\ref{fig:device-2d}(b), each single-lead quantum dot (SLQD) is positioned at the center of a rectangular configuration of four clusters, enabling both electron spin readout and initialization~\cite{hogg2023single}. Nuclear spins within a single cluster (Fig.~\ref{fig:device-2d}(c)) can be initialized and measured via electron-assisted quantum non-demolition (QND) readout. Additionally, micromagnets placed outside the array generate inhomogeneous magnetic field to improve spatial addressability of spin qubits in the clusters, and mitigates the complexity of fanout addressing lines by leveraging frequency multiplexing via a single microwave antenna. The antenna generates alternating magnetic fields $B_{ac}$ to drive both electron and nuclear spins. In Fig.~\ref{fig:device-2d}(d), the electron spins on two adjacent clusters are coupled by the exchange interaction, which is tunable through utilizing top gates (TGs), or by quantum dot (QD)-mediated superexchange interaction (discussed in Supplementary Materials Sec.~\ref{sec:QD}~\cite{supp})~\cite{Srinivasa2015,Malinowski2018,Baart2017,Munia2023,ZhangSH2025}. Multiple cluster arrays can be interconnected with the aid of moving QDs~\cite{Huang2013,zhao_doppler_2016,noiri_shuttling-based_2022,Langrock2023,xueR2024,de_smet2025} or cQED cavities~\cite{Hu2012,Mi2018,borjans2020,Harvey-Collard2022,yu2023,dijkema2025} to achieve further scaling (discussed in Sec.~\ref{sec:discussion}), paving the way for large-scale quantum processors based on spin qubits. 

    \begin{figure*}[tbph!]
		\begin{center}
			\includegraphics[width=2\columnwidth]{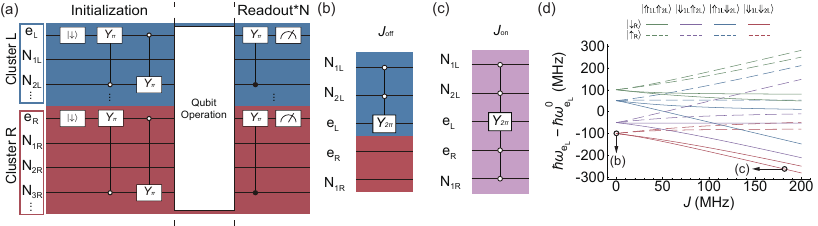}
		\end{center}
		\vspace{-0.50cm}
		\caption{Protocol of quantum processing. (a) Circuit for the control protocol includes qubit initialization, operation, and readout. Detailed sequences for the initialization and readout of a specific nuclear spin in each cluster are shown. Taking $\mathrm{N_{\rm{2L}}}$ $\mid\Downarrow\rangle$ initialization as an example~\cite{Thorvaldson2024,edlbauer2025,ZhangCH2025}, the electron is initially prepared to $\mid\downarrow\rangle$, then all ESR $\pi$ pulses involving $\mathrm{N_{\rm{2L}}}$ are applied to flip the corresponding electron spin; followed by the NMR $\pi$ pulse conditional on the electron spin being in $\mid\downarrow\rangle$. The three steps above constitute the initialization procedure for $\mathrm{N_{\rm{2L}}}$ to be $\mid\Downarrow\rangle$ state. The final nuclear spin readout, also implemented with the assistance of a bound electron, constitutes a QND measurement. This property enables repeated measurements to suppress SPAM errors. (b) Intra-cluster CZ operation based on a conditional 2$\pi$ ESR rotation with $J=J_{\rm{off}}$ [(b) in Figure (d)]. (c) Inter-cluster CZ operation based on a conditional 2$\pi$ ESR rotation with $J=J_{\rm{on}}$ [(c) in Figure (d)]. (d) The ESR frequency of the electron in the le cluster is shown as a function of the exchange interaction strength $J$. Here, a 2P-1P pair is taken as an example. Near $J=0$, the frequencies split into four groups depending on the nuclear spin states in the left cluster: $\mid\Downarrow_{\rm{1L}}\Downarrow_{\rm{2L}}\rangle$ (red), $\mid\Uparrow_{\rm{1L}}\Downarrow_{\rm{2L}}\rangle$ (purple), $\mid\Downarrow_{\rm{1L}}\Uparrow_{\rm{2L}}\rangle$ (blue), and $\mid\Uparrow_{\rm{1L}}\Uparrow_{\rm{2L}}\rangle$ (green). The solid curves correspond to the electron spin in the right cluster being $\mid\downarrow_{\rm{R}}\rangle$, while the dashed curves represent the electron spin being $\mid\uparrow_{\rm{R}}\rangle$. The splitting between frequencies with the same color and line style depends on the nuclear spin state in the cluster on the right. }
		\label{fig:control}
	\end{figure*}

    \subsection{System of donor-cluster arrays}
    
    We consider a system consists of an array of clusters (Fig.~\ref{fig:device-2d}(b)), each hosting a single electron and {multiple donor nuclei. The electrons in the neighboring clusters are coupled via the exchange interaction. In the presence of micromagnets and an applied magnetic field, a site-dependent magnetic field, $\mathbf{B}_{i}$ for the $i$-th cluster, is generated. 
    The Hamiltonian for the electron spins and nuclear spins in the cluster array is
	\begin{subequations}
		\begin{align}
			H &= \sum_i H_{i} + \sum_{\langle i,j\rangle}J_{i,j}(\epsilon)\mathbf{S}_{i}\cdot\mathbf{S}_{j}, \\
			H_{i} &= \gamma_{e}\mathbf{B}_{i}\cdot\mathbf{S}_{i} + \sum_{k}(\gamma_{n}\mathbf{B}_{i}\cdot \mathbf{I}_{i,k} + A_{i,k}\mathbf{S}_{i}\cdot\mathbf{I}_{i,k}),
		\end{align}
	\end{subequations}
	where $H_{i}$ is the single-cluster Hamiltonian for $i$-th cluster, and $J_{i,j}$ is the strength of the exchange interaction between nearest-neighbor electrons. 


    In the Hamiltonian, $H_{i}$ includes the Zeeman Hamiltonian and hyperfine (HF) Hamiltonian for the $i$-th cluster, which comprises $N_i$ donor nuclei (with distributed HF couplings $A_{i,k}$) sharing one electron.
    $\mathbf{S}$ and $\mathbf{I}_{i,k}$ are spin operators of the electron spin and the $k$-th nuclear spin within the $i$-th cluster, respectively. $\gamma_{n} = -17.41$ MHz/T and $\gamma_{e} = 27.97$ GHz/T are the gyromagnetic ratios of the nuclear spin and the electron spin, respectively. In the absence of time-dependent drives, only the static magnetic field $B_{i,0}$ remains, inducing Zeeman splittings of the electron spins and nuclear spins. Due to the differences in the spatial configuration of donors within the clusters, the HF values are distributed over a specific range. Typical values of $A_{i,k}$ in $H_i$ for P donors within a cluster range from hundreds of kHz to hundreds of MHz~\cite{Buch2013,Thorvaldson2024,edlbauer2025,ZhangCH2025,ZhangCH2026}.

    Due to the distribution of HF couplings, the resonance frequencies of different nuclear spins are distinct, and similarly for the electron spins. 
    In particular, when the exchange interaction is turned off, the energy of a given state is increased (or decreased) by $A_{i,k}/4$ when the electron spin and the $k$-th nuclear spin in the $i$-th cluster are paralleled (or anti-paralleled), resulting in the frequency-selective NMR (or ESR) for addressable control of the nuclear (or electron) spins.
    Although both single- and multi-qubit gates within a single cluster have been experimentally demonstrated~\cite{Thorvaldson2024,edlbauer2025}, along with recent realizations of inter-cluster multi-qubit gates through multiple pulses with a relatively weak exchange~\cite{edlbauer2025}, we further discuss the key challenges and requirements of universal gates for scaling up the cluster-based qubit architecture in the following section.


    \subsection{Quantum computing control protocol}\label{sec:protocol} 

    Fig.~\ref{fig:control}(a) shows the overall circuit for the control protocol of quantum information processing, including the initialization, gate operations, and readout process. To begin with, the tuning-up  of the ESR and NMR frequencies in the cluster array is performed, as detailed in the Sec.~\ref{sec: tuning-up} of the Supplementary Materials~\cite{supp}. Once the characterization of the system is done, we proceed to the initialization of a specific nuclear spin qubit~\cite{Thorvaldson2024,edlbauer2025,ZhangCH2025}: The electron spin is initialized by energy-selective tunneling, followed by a conditional ESR $\pi$ pulse that correlates the electron and nuclear spin. Afterwards, a final conditional NMR $\pi$ pulse completes the nuclear spin initialization. Then, the quantum information is processed afterwards,  built upon two elementary primitives: NMR-based single-qubit gates and ESR-based multi-qubit CZ gates (as detailed below). After qubit operations, high-fidelity QND readout of the nuclear spin qubit can be achieved by using an electron-assisted scheme~\cite{Thorvaldson2024,edlbauer2025,ZhangCH2025}. 

	\textit{Universal gate set.} Arbitrary unitary operations on qubits require a universal gate set, typically comprising at least one non-Clifford gate, one Clifford gate and one entangling two-qubit gate~\cite{Barenco1995universal_gate}. A typical universal gate set is \{$R_{y}(\pi/4)$, $S$, CNOT\}~\cite{Kitaev1997universal_gate}. In the cluster-based qubit scheme, arbitrary single-qubit gates based on NMR and multi-qubit CZ-type gates via conditional ESR can be natively implemented. Specifically, the $R_{y}(\pi/4)$ and $S$ gate can be realized via NMR-based operations. Alternatively, the $S$ gate can be realized by incorporating a virtual phase of $\pi/2$ into other pulses. Meanwhile, a multi-qubit CZ-type (and similarly a Toffoli) gate between nuclear spins can be implemented by the frequency-selective ESR conditional on the nuclear spin states.
    Consequently, a universal quantum gate set for nuclear spin qubits can be established, enabling arbitrary unitary operations.
    \textit{ESR spectrum for a cluster-based system.} Before discussing ESR-based multi-qubit gates, we first introduce the ESR spectrum using a 2P-1P pair as an illustrative example. For a 2P-1P pair with HF strengths [$A_{\rm{1L}}$, $A_{\rm{2L}}$, $A_{\rm{1R}}$] = [50, 150, 110] MHz, the ESR spectrum of the electron in the left cluster is plotted as a function of $J$ in Fig.~\ref{fig:control}(d). The magnetic field $B_{0}$ is set to 1.35 T. Due to the HF interaction, the ESR frequencies split into four groups near $J=0$, corresponding to the following nuclear spin states: $\mid\Downarrow_{\rm{1L}}\Downarrow_{\rm{2L}}\rangle$ (red), $\mid\Uparrow_{\rm{1L}}\Downarrow_{\rm{2L}}\rangle$ (purple), $\mid\Downarrow_{\rm{1L}}\Uparrow_{\rm{2L}}\rangle$ (blue), and $\mid\Uparrow_{\rm{1L}}\Uparrow_{\rm{2L}}\rangle$ (green), respectively. As the exchange coupling increases from zero, each of these four groups of ESR frequencies further split into solid and dashed lines, corresponding to the $\mid\downarrow_{\rm{R}}\rangle$ and $\mid\uparrow_{\rm{R}}\rangle$ states of the right electron, respectively. The frequency shift, conditioned on the neighboring nuclear spin states, increases with $J$~\cite{Kranz2023,stemp2025}, which is consistent with the analytical results (see Supplementary Materials Sec.~\ref{sec:analytical})~\cite{supp}. When $J$ is sufficiently large, these lines are further divided into two lines, corresponding to the adjacent nuclear spin states being in the $\mid\Uparrow_{\rm{1R}}\rangle$ and $\mid\Downarrow_{\rm{1R}}\rangle$ states.

    \textit{Intra-cluster multi-qubit gates using ESR.} During intra-cluster operations, the exchange interaction is turned off to suppress crosstalk. The intra-cluster CZ-type gate is implemented by imparting a $\pi$ phase conditional on specific states of nuclear spins within a cluster. For example, to implement the intra-cluster CZ gate (see Fig.~\ref{fig:control}(b)), a $2\pi$ ESR pulse is applied at the frequency corresponding to point B in Fig.~\ref{fig:control}(d), leading to the accumulation of a $\pi$ geometric phase on the $\mid\Downarrow_{\rm{1L}}\Downarrow_{\rm{2L}}\rangle$ state~\cite{Madzik2022,Thorvaldson2024}. Moreover, the intra-cluster CNOT gate can be implemented by applying $H$ gates to the target qubit before and after the CZ gate~\cite{Madzik2022}.
    
    \textit{Inter-cluster multi-qubit gates using ESR.} The implementation of inter-cluster multi-qubit gates is essential for the scalability of the cluster array. The inter-cluster multi-qubit operations between adjacent clusters are mediated by exchange-coupled electron spins. As shown in Fig.~\ref{fig:control}(d), the activated exchange coupling causes the ESR frequency to depend on the nuclear spin configuration in the adjacent cluster. Consequently, a 2$\pi$-rotation on the electron spin can be executed conditional on the joint state of nuclear spins across two neighboring clusters. As exemplified in Fig.~\ref{fig:control}(c), an ESR $2\pi$ $Y$ rotation on the left electron spin is applied conditional on the state $\mid\Downarrow_{\rm{1L}}\Downarrow_{\rm{2L}}\downarrow_{\rm{R}}\Downarrow_{\rm{1R}}\rangle$, resulting in an inter-cluster multi-qubit CZ gate. The ESR frequency corresponds to (c) in Fig.~\ref{fig:control}(d). A moderate $J$ underpins the direct implementation of inter-cluster multi-qubit gates. Similar to the intra-cluster case, inter-cluster multi-qubit Toffoli gates can be realized. The overhead of these multi-qubit gates is discussed in Supplementary Materials Sec.~\ref{sec:overhead}~\cite{supp}. In the following, this implementation scheme is referred to as E-TCMG (ESR-based two-cluster multi-qubit gate).

    \begin{figure}[tbp]
		\begin{center}
			\includegraphics[width=0.99\columnwidth]{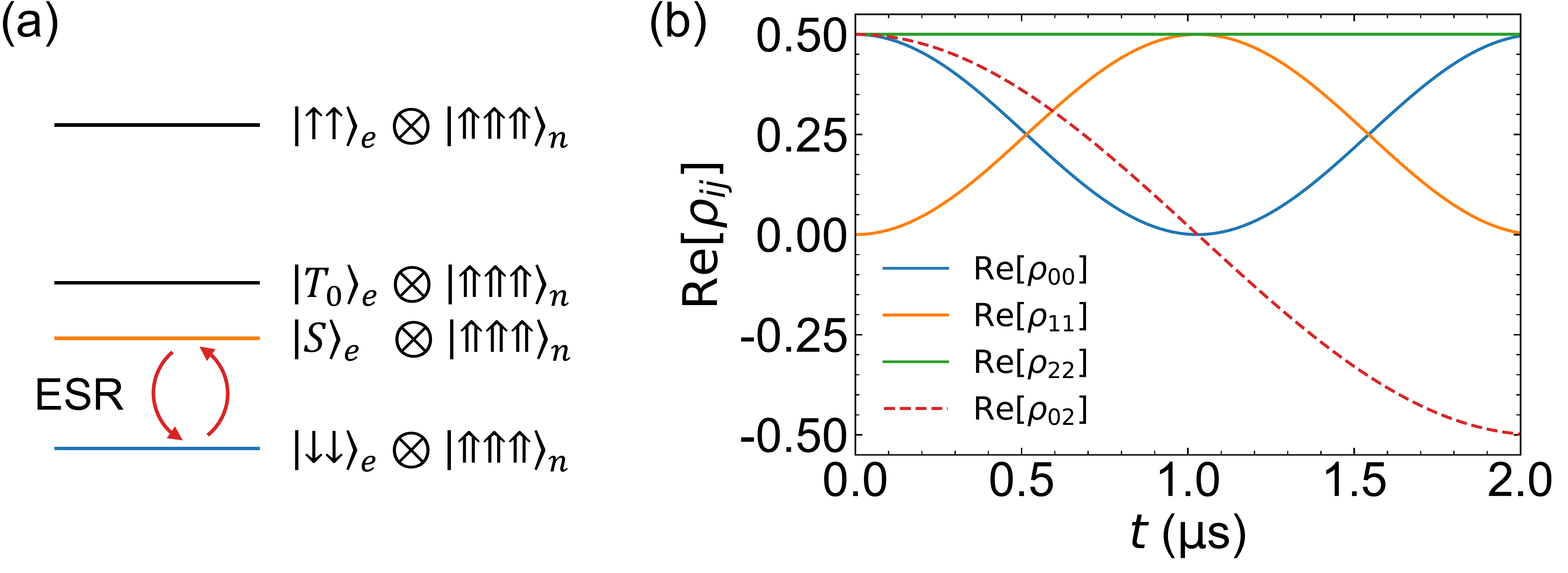}
		\end{center}
		\vspace{-0.50cm}
		\caption{Singlet-triplet-based inter-cluster CZ gate. (a) The energy levels for two electron spins in the strong coupling regime. (b) The time evolution of the real part of the density matrix elements $\rho_{00}$, $\rho_{11}$, $\rho_{22}$ and $\rho_{02}$ in a 1P-2P system. The diagonal terms (solid lines) and $\mathrm{Re}[\rho_{02}]$ (dashed line) are plotted as a function of evolution time $t$.}
		\label{fig:ST0}
	\end{figure}
    
    \textit{Multi-qubit gate based on two-electron singlet-triplet states.}
In the regime where the exchange coupling is strong compared with the energy difference between the two adjacent electron spins, the anti-parallel electron spin states hybridize to form the spin singlet $\vert S \rangle_e$ and triplet $\vert T_0 \rangle_e$ states, in addition to the polarized $\mid \uparrow\uparrow \rangle_e$ and $\mid \downarrow\downarrow \rangle_e$ states (Fig.~\ref{fig:ST0}(a)). In this case, it is difficult to distinguish each individual electron spins and crosstalk errors could be induced. The realization of the nuclear-nuclear CZ gate hinges on the accumulation of a geometric phase of $\pi$ on the target nuclear spin state via a $2\pi$ rotation of the electron spin. In the strong $J$ regime, a rotation between the $\mid \downarrow\downarrow \rangle_e$ and $\vert S \rangle_e$ state can also be achieved via ESR, so that the inter-cluster multi-qubit CZ gate on nuclear spins can still be realized. This implementation scheme is termed ST-TCMG (singlet-triplet-based TCMG).

To validate this approach, we perform numerical simulations of the dynamical process for a 1P-2P system with HF interactions $[A_{\rm{1L}}, A_{\rm{1R}}, A_{\rm{2R}}] = [110, 50, 150]$~MHz. We define the states $\vert 0 \rangle = {\mid \downarrow\downarrow \rangle_e} \otimes \mid \Downarrow\Downarrow\Downarrow \rangle_n$, $\vert 1 \rangle = \vert S \rangle_e \otimes \mid \Downarrow\Downarrow\Downarrow \rangle_n$, and $\vert 2 \rangle = {\mid \downarrow\downarrow \rangle_e} \otimes \mid \Downarrow\Downarrow\Uparrow \rangle_n$. The system is initialized in the $\vert 0 \rangle$ state, and an ESR pulse is applied at frequency $\hbar\omega = E_{\vert 1 \rangle} - E_{\vert 0 \rangle}$. As shown in Fig.~\ref{fig:ST0}(b), under ESR driving, the system evolves from state $\vert 0 \rangle$ to $\vert 1 \rangle$ and subsequently returns to $\vert 0 \rangle$, ensuring that all populations return to their initial values. In contrast, $\mathrm{Re}[\rho_{02}]$ (representing the phase between states $\vert 0 \rangle$ and $\vert 2 \rangle$) acquires a sign change, indicating the accumulation of a $\pi$ phase. This confirms that the singlet-triplet-based ESR operation successfully implements the conditional phase gate.

\added{Compared to the E-TCMG scheme, the singlet-triplet-based CZ gate avoids the difficulty of distinguishing individual electron spins in the strong coupling regime, thereby relaxing the parameter requirements for high-fidelity inter-cluster multi-qubit gates.}
E-TCMG and ST-TCMG together significantly expands the acceptable parameter regions for exchange interaction strength and HF coupling differences, offering higher parameter tolerance and enhanced feasibility for the cluster-based spin qubit scheme.

\subsection{Crosstalk analysis for qubit operations}\label{sec:addressability} 
    Since the addressability of quantum operations for both the nuclear and electron spins in the donor-cluster array relies on the frequency selectivity, mitigating frequency crowding is crucial for addressable high-fidelity quantum operations and thus essential for scalability of the quantum computing architectures. In particular, sufficient frequency separations are critical for ESR-based multi-qubit gates, as the fast ESR operations require larger frequency separations to suppress crosstalk errors. Due to the stochastic placement of donors within clusters, HF interaction strengths can vary significantly across different donor configurations and donor numbers. The distribution in HF interaction strengths naturally results in intrinsic frequency differentiation. 

    \added{The crosstalk errors during NMR, intra-cluster, and inter-cluster ESR operations are analyzed (see Supplementary Materials Sec.~\ref{sec:crosstalk_result}~\cite{supp} for details).
    While NMR introduces limited crosstalk due to its weak driving strength, the dominant errors originate from ESR operations. For intra-cluster ESR, crosstalk error rate can be suppressed below 5\% provided that the residual exchange coupling is weak ($J_{\mathrm{off}}<10$ MHz) and the minimum HF difference is sufficiently large $\Delta A>3$ MHz). In contrast, achieving the same error threshold for the inter-cluster ESR operation (E-TCMG) requires a strong activated exchange interaction ($J_{\mathrm{on}}> 50$ MHz) along with a corresponding $\Delta A > 25$ MHz.
    Both the crosstalk error rate and the viable parameter range can be further optimized by employing weaker driving strengths or additional magnetic field gradients. Furthermore, initializing and idling the relevant nuclear spins offers another mitigation strategy, which is expected to relax the requirements on $\Delta A$ and $J$.
    }
    
    To further mitigate crosstalk errors arising from the direct implementation of inter-cluster CZ gates, two types of indirect multi-qubit gates implemented via auxiliary operations can be employed: ESR-assisted and NMR-assisted multi-qubit gates, which are referred to as EA-TCMG (ESR-assisted TCMG) and NA-TCMG (NMR-assisted TCMG). Their implementations and performance are detailed in Supplementary Materials Sec.~\ref{sec:assist}~\cite{supp}. With a weak exchange interaction ($\sim$ 10 MHz), the EA-TCMG-based CZ gate relaxes the requirement for the HF difference $\Delta A$ compared to the directly implemented CZ gate. Meanwhile, for the NA-TCMG, leveraging the high tunability of $J$, both the required $\Delta A$ and the gate infidelity are reduced.
    
     In the next section, we evaluate the performance of multi-qubit gates and identify the feasible experimental parameter range through analytical estimations and numerical simulations. We also study the performance of the inter-cluster CZ gate based on the two-electron spin singlet and triplet states, which further relax the requirements on parameters. Moreover, we show the NA-TCMG-based CZ gate exhibits significantly lower crosstalk errors compared to the direct implementation in multi-cluster configurations.
     
    \section{Performance of multi-qubit gates between spin qubits in clusters} \label{sec: gate_performance}
    \subsection{Fidelity of multi-qubit gates}\label{sec: anal_fid}
        \begin{figure*}[hbt!]
		\begin{center}
			\includegraphics[width=2\columnwidth]{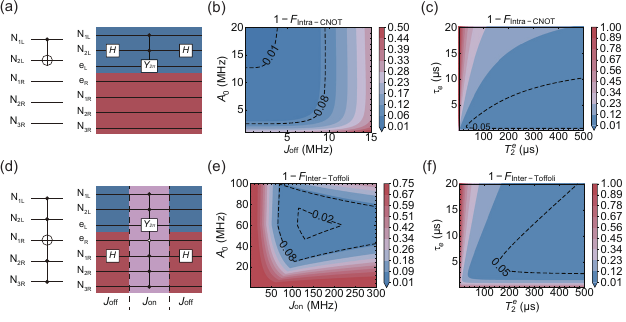}
		\end{center}
		\vspace{-0.50cm}
		\caption{Intra-cluster CNOT gate and inter-cluster Toffoli-type multi-qubit gate between nuclear spin qubits. (a) The intra-cluster CNOT gate with $\mathrm{N}_{\mathrm{1L}}$ as the control qubit and $\mathrm{N}_{\mathrm{2L}}$ as the target qubit and its compiled circuit. The CNOT gate is composed of an ESR-based CZ gate and two single-qubit gates applied to $\mathrm{N}_{\mathrm{2L}}$. (b) The infidelity $1-F_{\rm{Intra-CNOT}}$ of the intra-cluster CNOT-type gate is plotted as a function of the residual exchange interaction $J_{\rm{off}}$ and the hyperfine interaction $A_{0}$. High fidelity requires suppressing $J_{\rm{off}}$. (c) $1-F_{\rm{Intra-CNOT}}$ versus the decoherence time $T_{2}^{e}$ and the 2$\pi$-rotation time $\tau_{e}$ of the electron spin. (d) The inter-cluster five-qubit Toffoli gate with $\mathrm{N}_{\mathrm{1R}}$ as the target qubit and the other four nuclear spins as control qubits, along with its compiled circuit. The Toffoli gate is composed of an ESR-based inter-cluster CCCCZ gate and two single-qubit gates applied to $\mathrm{N}_{\mathrm{1R}}$. (e) The infidelity $1-F_{\rm{Inter-Toffoli}}$ of the inter-cluster Toffoli-type gate versus the activated exchange interaction $J_{\rm{on}}$ and $A_{0}$. The lower fidelity compared to (b) reflects the higher crosstalk of the inter-cluster multi-qubit gate. (f) $1-F_{\rm{Inter-Toffoli}}$ versus a function of the decoherence time $T_{2}^{e}$ and the 2$\pi$-rotation time $\tau_{e}$ of the electron spin.}
		\label{fig:gate_ct}
	\end{figure*}
    In this section, we present the ESR-based nulcear-spin multi-qubit gate scheme and characterize gate performance metrics within our cluster-based qubit scheme. Since the Toffoli gate includes ESR and NMR operations, we use it as a representative multi-qubit gate. Without loss of generality, we take the 2P-3P cluster pair as an example, as shown in Fig.~\ref{fig:device-2d}(d), and the HF interaction strengths in the cluster pair are given by $[A_{\mathrm{1L}}, A_{\mathrm{2L}}, A_{\mathrm{1R}}, A_{\mathrm{2R}}, A_{\mathrm{3R}}] = [A_0, 170, 60, 120, 230]$ MHz for intra-cluster CNOT gate and $[A_{\rm{1L}}, A_{\rm{2L}}, A_{\rm{1R}}, A_{\rm{2R}}, A_{\rm{3R}}] = [60, 170, A_{0}, 120, 230]$ MHz for inter-cluster Toffoli gate. 
    We use $A_{0}$ as the variable for simplicity instead of the minimal HF difference $\Delta A$. 
    Errors induced by crosstalk and single-qubit decoherence are considered (see Supplementary Materials sec.~\ref{sec:error_anal} for details~\cite{supp}). The dephasing time of the nuclear spin is set to be 40 ms~\cite{edlbauer2025}. The strength of the time-dependent NMR driving is set to be 0.6 mT~\cite{ZhangCH2025}, corresponding to a Rabi frequency of approximately 30.36 kHz for a single P donor. Since the electron spin only undergoes a $2\pi$ rotation, we set the electron spin decoherence time to be 400 $\upmu$s, corresponding to the spin Rabi dephasing time~\cite{tyryshkin2012}. The Rabi frequency of intra-cluster (or inter-cluster) ESR operations is assumed to be 0.5 MHz (or 0.3 MHz)~\cite{Madzik2022}.
    
    
    We first evaluate the performance of intra-cluster CNOT-qubit gates. Figure \ref{fig:gate_ct}(a) shows the compiled circuit for a nuclear-nuclear CNOT gate within the 2P cluster (with $\rm{N_{2L}}$ as the target qubit). The CNOT gate is realized by applying Hadamard $H$ gates to the target nuclear spin qubit before and after the CZ gate. To give a conservative performance estimation, we calculate the infidelity of intra-cluster CNOT gates with the most severe crosstalk in the ESR operations. In Fig.~\ref{fig:gate_ct}(b), the dependence of the CNOT gate infidelity $1-F_{\rm{Intra-CNOT}}$ on $J_{\rm{off}}$ and $\Delta A$ is similar to that of the crosstalk error of intra-cluster ESR operations (Supplementary Materials Fig.~\ref{fig:control-Addressibility}(b)~\cite{supp}). High-fidelity ($>99\%$) intra-cluster CNOT gate is achievable. Figure \ref{fig:gate_ct}(c) shows the effect of single-qubit decoherence and ESR operation time $\tau_{e}$, where the black dashed line corresponds to an infidelity of $1-F_{\rm{Intra-CNOT}}<0.05$ ($J_{\rm{off}} = 3$ MHz and $A_{0}= 15$ MHz).
    The upper bound of the ESR operation time $\tau_{e}$ is determined by the electron decoherence time $T_{2}^{e}$, while the lower bound is limited by the minimum frequency detuning required to avoid driving-induced crosstalk errors. 

    Next, we evaluate the performance of inter-cluster multi-qubit gates (E-TCMG). Figure \ref{fig:gate_ct}(d) shows the compiled circuit for a five-qubit Toffoli gate between the 2P and 3P clusters with $\rm{N_{1R}}$ as the target qubit. Similarly, to assess the performance in the worst case, only the five-qubit CZ gate with the minimum difference between ESR frequencies is considered, and the infidelity $1-F_{\rm{Inter-Toffoli}}$ of the inter-cluster Toffoli-type gate is calculated, as shown in Fig.~\ref{fig:gate_ct}(e). A region with low infidelity ($<0.08$) is achieved for HF strength ranging from 30 MHz to 100 MHz and the exchange interaction above 80 MHz. The fidelity of the inter-cluster Toffoli gate can further exceed 98\% in certain parameter regimes. Figure \ref{fig:gate_ct}(f) shows the infidelity $1-F_{\rm{Inter-Toffoli}}$ as a function of the decoherence time $T_{2}^{e}$ and the operation time $\tau_{e}$ of the electron spin. Compared to the results of intra-cluster CNOT gates, achieving a low infidelity ($<0.05$) requires a longer $T_{2}^{e}$ and permits a shorter feasible range of $\tau_{e}$. This is primarily because crosstalk errors are more severe when performing inter-cluster Toffoli gates, further emphasizing the importance of crosstalk suppression.
    \begin{figure*}[htbp]
		\begin{center}
			\includegraphics[width=1.9\columnwidth]{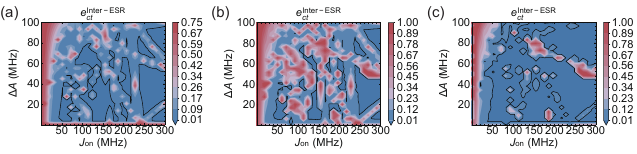}
		\end{center}
		\vspace{-0.50cm}
		\caption{Crosstalk error of the inter-cluster ESR operation on spin singlet-triplet mixed state for a 2P-1P system. The black lines in figures indicate the region with $e_{ct}^{\rm{Inter-ESR}}<1\%$. (a) and (b) The infidelity $e_{ct}^{\rm{Inter-ESR}}$ of the inter-cluster ESR operation is shown as a function of $J_{\rm{on}}$ and $\Delta A$, considering the singlet–triplet mixed state of the electron spins. Panel a is obtained by analytical estimation, while panel b is calculated numerically. (c) The infidelity $e_{ct}^{\rm{Inter-ESR}}$ calculated numerically when the final states of the electron spins are neglected by tracing out their degrees of freedom. As a result, the fidelity and the range of viable parameters are improved.}
		\label{fig:ct_nuc}
	\end{figure*}

    \begin{figure*}[htbp]
		\begin{center}
			\includegraphics[width=1.9\columnwidth]{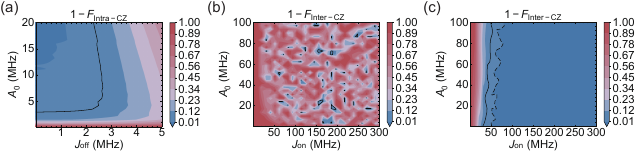}
		\end{center}
		\vspace{-0.50cm}
		\caption{Infidelity of the multi-qubit CZ gate for a 2P-3P system. The black solid lines in figures indicate the region with $1-F_{\rm{Inter-CZ}}<5\%$. (a) The infidelity $1-F_{\rm{Intra-CZ}}$ of the intra-cluster two-qubit CZ gate is plotted as a function of $J_{\rm{off}}$ and $A_0$. (b) The infidelity $1-F_{\rm{Inter-CZ}}$ of the direct-implemented inter-cluster CZ gate versus $J_{\rm{on}}$ and $A_0$, considering spin singlet-triplet mixed state of electrons. (c) The infidelity $1-F_{\rm{Inter-CZ}}$ of the NA-TCMG-based inter-cluster CZ gate versus $J_{\rm{on}}$ and $A_0$, considering spin singlet-triplet mixed state of electrons. The black dashed lines indicate the region with $1-F_{\rm{Inter-CZ}}<1\%$. The NA-TCMG-based implementation suppresses the crosstalk errors, achieving a broad high-fidelity region across a wide range of $J_{\rm{on}}$ and $A_0$.}
		\label{fig:cz_num}
	\end{figure*}
    
    To evaluate the practical ESR operation performance, we calculated the inter-cluster ESR operation infidelities numerically for a 1P-2P cluster pair (see Supplementary Materials Sec.~\ref{sec:numerical-results}~\cite{supp}). In the numerical simulations, we considered the minimal frequency separation, accounting for also the accidental degeneracies among ESR frequencies. In region where $J_{\rm{on}}$ is much stronger than $\Delta A$, the infidelity is significantly higher. This is because the eigenstates of two electron spins are spin singlet-triplet states due to the strong coupling between electron spins. In this case, the ESR frequencies of the two electron spins are nearly degenerate, which induces crosstalk errors during the ESR operation.

    \subsection{ST-TCMG and numerical results}

    To relax the system parameter requirements for implementing inter-cluster CZ gates, we propose leveraging the two-electron spin singlet and triplet states. Since implementing a nuclear-nuclear multi-qubit gate only requires a $2\pi$ rotation of the electron spins to accumulate a geometric $\pi$ phase, performing the rotation between the two-electron $\mid\downarrow\downarrow\rangle$ state and $|S\rangle$ state can also result in the effective CZ gate (Fig.~\ref{fig:ST0}). Allowing the ESR operation on in the two-electron spin subspace, we calculate the ESR operation infidelities both analytically and numerically. For comparison, the corresponding results are shown in Fig.~\ref{fig:ct_nuc}(a) and (b), respectively. Additionally, to evaluate the impact of the electron spin on the fidelity calculation, Figure \ref{fig:ct_nuc}(c) plots the results with the electron spin traced out. The conditions for the figures are exactly the same. The black-line-indicated regions correspond to an infidelity below 1\%. Utilizing the two-electron spin subspaces significantly expands the acceptable parameter regions, offering higher parameter tolerance and feasibility for the cluster-based spin qubit scheme. When $J$ is sufficiently strong, the range of $\Delta A$ for achieving high fidelity is wider.
    
	
	In the calculation of crosstalk errors in Fig.~\ref{fig:ct_nuc}(a) and (b), both electron and nuclear spin states are taken into account. However, in the cluster-based nuclear spin qubit scheme, electron spins act as ancilla rather than data qubits, so that the electron spin states only indirectly influence the operations of nuclear spins. Hence, the practical fidelity of the ESR operation is expected to be higher than the results shown in Fig~\ref{fig:ct_nuc}(b), which represent the lower bound of the fidelity. To evaluate the practical performance of the ESR operation, an upper bound can be obtained by calculating the fidelity while tracing out the electron spin states. Figure \ref{fig:ct_nuc}(c) shows the multi-qubit gate fidelity when tracing out the electron spin states. As expected, $e_{ct}^{\rm{Inter-ESR}}$ is lower than that of Fig.~\ref{fig:ct_nuc}(b). In addition, the feasible region of parameters is also significantly expanded. As long as the impact of the electron spin can be eliminated, the fidelity of nuclear spin qubits can be significantly enhanced.

    Next, we provide a comprehensive numerical evaluation of the infidelity of intra- and inter-cluster multi-qubit gates, accounting for accidental degeneracies and single-qubit dephasing. Again, the worst-case scenarios are considered. Since the ESR operation constitutes the main performance bottleneck, we focus on the ESR-based CZ gate in a 2P-3P cluster pair for simplicity, rather than the CNOT gate. The parameters used in this section are consistent with those in Sec.~\ref{sec: anal_fid}. In Figs.~\ref{fig:cz_num}(a) and (c), the electron spin dephasing noise strength is set to 10 kHz with an ESR 2$\pi$-rotation duration of 2 $\upmu$s. In contrast, for Fig.~\ref{fig:cz_num}(b), the dephasing noise strength is reduced to 4 kHz and the 2$\pi$-rotation duration is 10 $\upmu$s to further suppress crosstalk errors. 
    
    Figure \ref{fig:cz_num}(a) shows the infidelity $1-F_{\rm{Intra-CZ}}$ of the intra-cluster CZ gate is plotted as a function of $J_{\rm{off}}$ and $A_0$. The region where $1-F_{\rm{Intra-CZ}}<5\%$ is indicated by the black solid line. The minimum HF difference $\Delta A$ should be at least 6 MHz, and the residual exchange interaction should be below 1.8 MHz.
    In Fig.~\ref{fig:cz_num}(b), the infidelity $1-F_{\rm{Inter-CZ}}$ of the inter-cluster CZ gate is plotted as a function of $J_{\rm{on}}$ and $A_0$. Due to frequent accidental degeneracies in the ESR frequencies, the parameter regions with infidelity $<5\%$ (indicated by the black lines) are fragmented and small, demanding high tunability of the exchange interaction. 
    
    As discussed in Sec.~\ref{sec:protocol}, the NA-TCMG-based CZ gate relaxes these parameter constraints. Figure \ref{fig:cz_num}(c) shows the infidelity $1-F_{\rm{Inter-CZ}}$ for the NA-TCMG-based CZ gate. The parameter region where an infidelity below $5\%$ for the NA-TCMG-based CZ gate is significantly larger than that for the directly implemented CZ gate. Moreover, the fidelity can be higher than 99\%, \added{exceeding the FT threshold of the surface code QEC (inter-cluster two-qubit CZ gate fidelity also exceeds the FT threshold, as estimated in Supplementary Materials Sec.~\ref{sec:overhead}~\cite{supp})}. Notably, this is achieved even though the noise strength considered for the NA-TCMG-based CZ gate is stronger than that for the direct-implemented CZ gate. The suppression of crosstalk errors under the NA-TCMG-based scheme significantly improves the inter-cluster CZ gate fidelity. Although the NA-TCMG-based CZ gate requires additional NMR operations that suffer from dephasing, the dephasing-induced biased error can be further suppressed by utilizing the bias-tailored QEC~\cite{ataides2021,Roffe2023}.

    \subsection{Multi-qubit gate protocols}\label{sec:TCMG}
    \begin{figure}[htbp]
    		\begin{center}
    			\includegraphics[width=0.98\columnwidth]{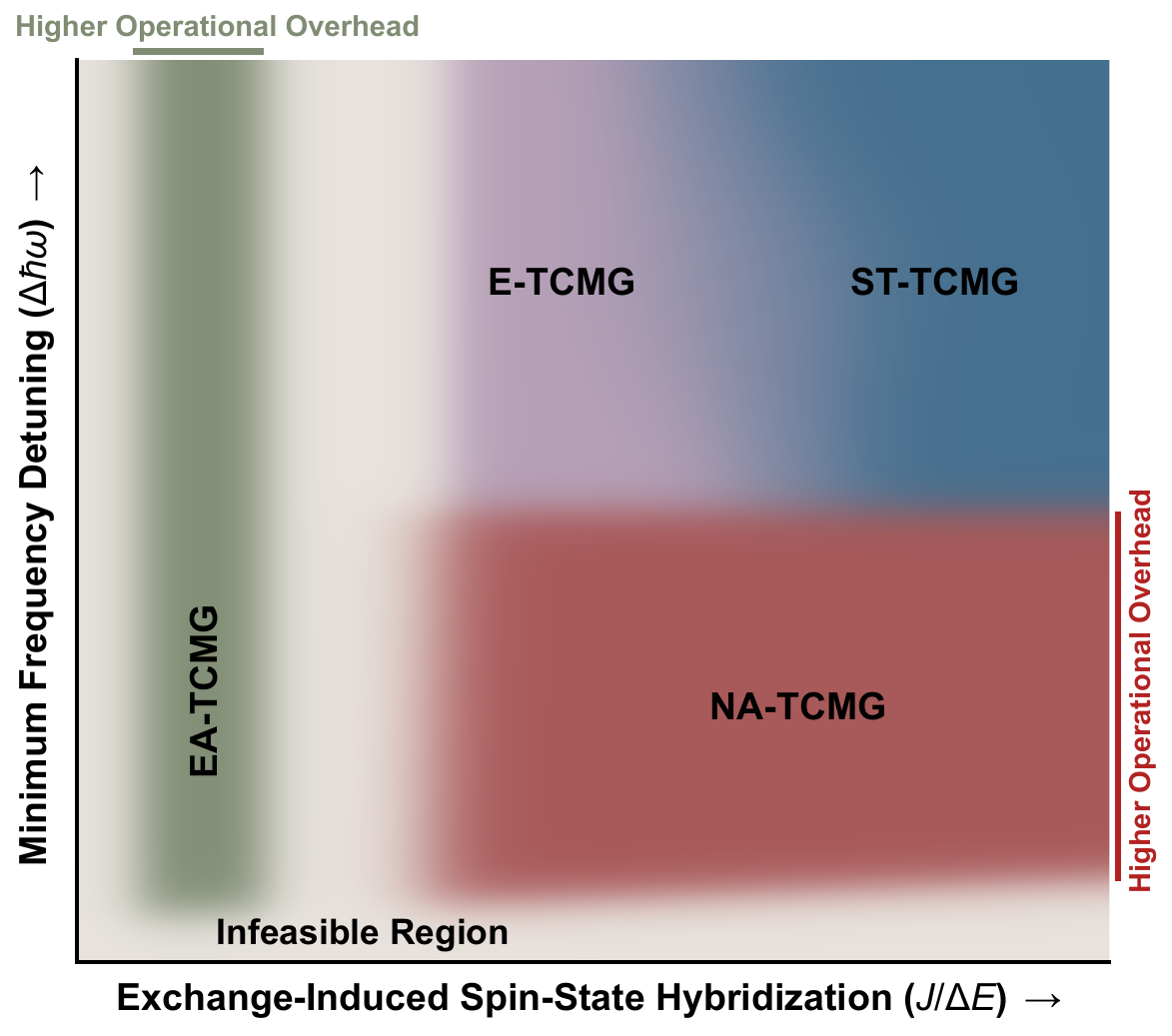}
    		\end{center}
    		\vspace{-0.50cm}
    		\caption{Recommended two-cluster multi-qubit gates (TCMGs) schemes for different parameter regimes. The diagram summarizes the preferred applicability regimes of different TCMG implementations as function of the exchange-induced spin-state hybridization ($J/\Delta E$) and the minimum frequency detuning $\Delta\hbar\omega$ between all ESR frequencies, where $J$ represents the exchange interaction strength and $\Delta E$ denotes the energy detuning between the two corresponding electron spin states. The four labeled regimes correspond to ESR-based TCMG (E-TCMG), singlet-triplet-based TCMG (ST-TCMG), ESR-assisted TCMG (EA-TCMG), and NMR-assisted TCMG (NA-TCMG). The shaded infeasible regions indicate parameter ranges in which none of these schemes is favorable. The red and green line outside the phase diagram highlight increasing operational overhead using NA/EA-TCMG.}
    		\label{fig:TCMG}
    	\end{figure}
    
    \added{In this paper, we have introduced four schemes for realizing inter-cluster multi-qubit gates, termed TCMGs. TCMG can be implemented directly and natively through either an E-TCMG or a ST-TCMG. However, in the presence of frequency crowding, performing these operations directly may induce strong crosstalk errors. To mitigate these errors, the EA-TCMG and NA-TCMG schemes, described in Supplementary Materials Sec.~\ref{sec:assist}~\cite{supp}, are introduced. Consequently, the choice of which TCMG scheme to implement should be made based on the specific experimental parameters and the requirements of the target quantum gate.}
    
    \added{The applicability regimes of the TCMG schemes are summarized in the diagram of Fig.~\ref{fig:TCMG}, which is plotted against the spin-state hybridization ($J/\Delta E$) and the minimum detuning $\Delta\hbar\omega$ between each two of all ESR frequencies. For sufficiently large frequency detuning, where crosstalk errors are suppressed, TCMGs can be implemented directly, with E-TCMG or ST-TCMG selected according to the eigenstate determined by the exchange interaction strength. By contrast, for smaller frequency detuning, indirect implementations assisted by additional operations are required. In the weak-exchange regime, where the ESR frequency depends on the spin state of the neighboring electron but remains independent of the nuclear-spin state of the neighboring cluster, EA-TCMG is more favorable. In the strong-exchange regime, NA-TCMG is more effective in suppressing crosstalk arising from frequency crowding. When the exchange interaction is too weak or the ESR frequency detuning is too small, none of the four schemes is viable, defining the infeasible region. In practice, the appropriate TCMG protocol should be selected based on the specific experimental conditions.} 

	\section{Discussion}\label{sec:discussion}
    There are several interesting features of our proposed cluster-based quantum computing architecture. First, certain randomness in both the position and the number of donors within each cluster is acceptable, which reduces the challenges associated with the deterministic placement of donors and thereby facilitates the fabrication process. Moreover, inherent randomness in the HF interaction strengths of donors results in a frequency distribution of spin qubits, which suppresses crosstalk errors and enables the addressability of the donor nuclear spins. Therefore, this donor cluster-based architecture harnesses rather than eliminates fabrication randomness, greatly improving fabrication efficiency.    

    \subsection{Scalability of the donor-cluster array}
    To realize large-scale quantum processors, it is essential to evaluate the scalability of the donor-cluster array. In particular, the addressability of qubits becomes increasingly challenging as the system size grows. In this section, we discuss the feasible number of qubits and potential strategies for preserving qubit addressability in large-scale arrays. 
    As discussed in Supplementary Materials Sec.~\ref{sec: number}~\cite{supp}, by simulating with 1000 samplings, the feasible number of donors within a cluster ranges from 2 to 11. The average number is 4.3 (1.5). In such case, the HF interaction strength distributions generally span a range from hundreds of kHz to hundreds of MHz~\cite{Buch2013,Thorvaldson2024,edlbauer2025,ZhangCH2025}. For larger-scale quantum systems (containing hundreds of qubits), HF interactions alone are insufficient to ensure frequency distinguishability. In this case, we incorporate micromagnets to induce magnetic field gradients, enabling the addressability of clusters~\cite{WangH2026}.

    Given the state-of-the-art magnetic field gradient ($\approx 0.45$ mT/nm~\cite{WangH2026}) provided by the micromagnets and the bandwidth constraints of a single microwave antenna, the practical number of qubits addressable in a single array is limited to the order of hundreds. Assuming a magnetic field gradient distributed laterally across the cluster array with an inter-column spacing of 13 nm, and this gradient is maintained over a range of 230 nm, the array could accommodate a maximum of about 120 donors~\cite{WangH2026}. Frequency crowding can be mitigated through the control protocol design to suppress crosstalk, as discussed in Ref.~\cite{ZhangCH2025}. In addition, affected qubits may be initialized but left idle during operations to avoid the crosstalk issue. Alternatively, local electrical manipulation schemes, such as electric-dipole spin resonance (EDSR)~\cite{Golovach2006,Tokura2006,Tosi2017}, could be employed to circumvent the frequency crowding issue. These strategies offer promising pathways for increasing the total number of integrated qubits in a donor-cluster array.

    \subsection{Compatibility with QEC code}\label{sec:QEC}
    
    	\begin{figure}[htbp!]
		\begin{center}
			\includegraphics[width=0.95\columnwidth]{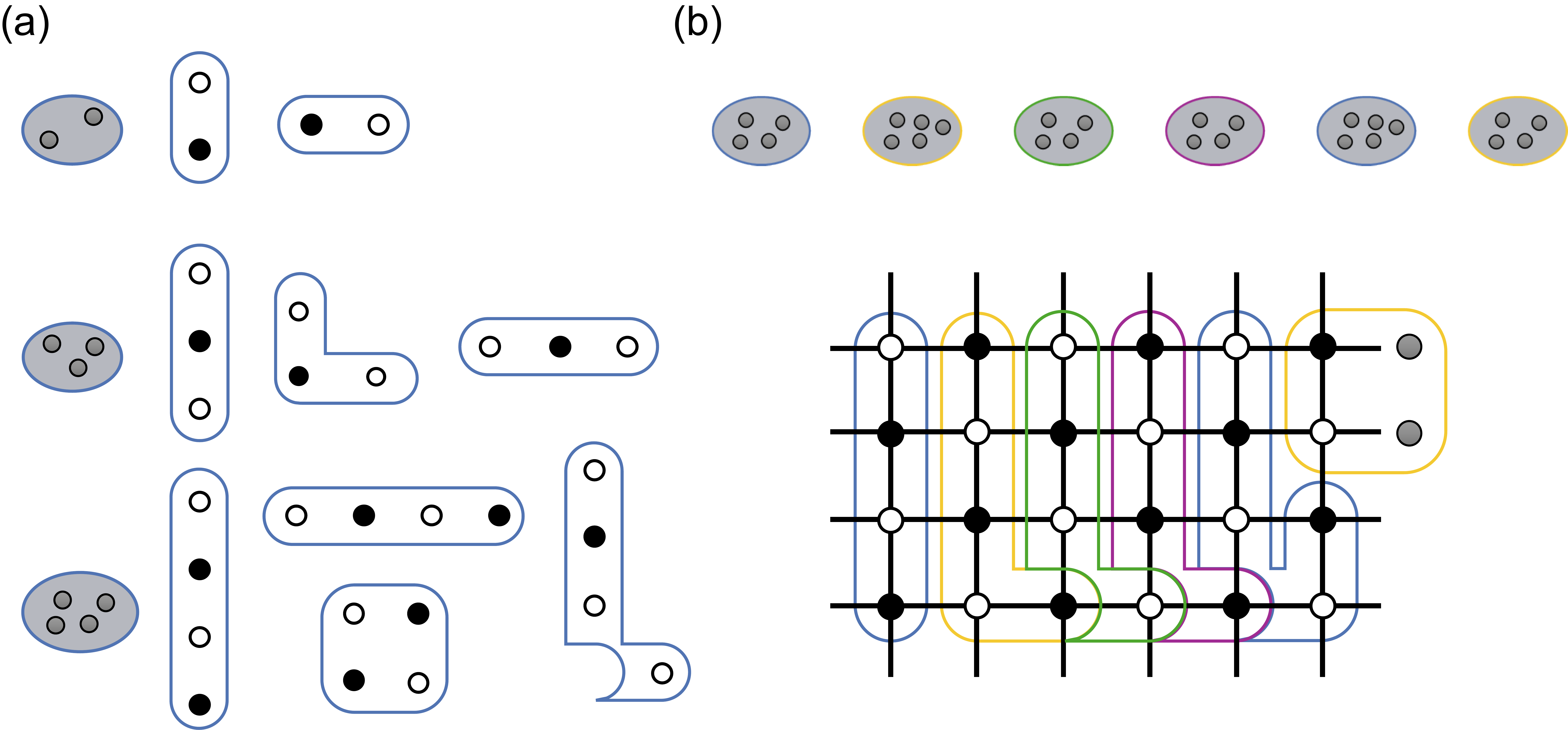}
		\end{center}
		\vspace{-0.50cm}
		\caption{QEC in cluster-based qubit systems. Gray dots represent nuclear spin qubits within clusters. White dots indicate data qubits, while black dots represent ancillary qubits. (a) The mapping between clusters and parts of the 2D toric code is shown. The shape of the mapped modules constructed by qubits within a cluster can be modified arbitrarily due to the all-to-all connectivity among them. (b) A 2D [[12,2,3]] XZZX toric code is constructed using a 1D cluster chain. Vertical lines represent Z-type Pauli checks, while horizontal lines indicate X-type Pauli checks.}
		\label{fig:QEC}
	\end{figure}

    The cluster-based architecture offers distinct advantages for QEC, thanks to its local all-to-all connectivity and capacity for low-overhead multi-qubit gates. These features enable the optimized implementation of QEC codes, such as the surface code, with each cluster serving as a reconfigurable module.
    As illustrated in Fig.~\ref{fig:QEC}(a), a 4P-cluster can be arranged into various configurations, including horizontal chain, vertical chain, square lattice, and other topological arrangements. This characteristic not only ensures the flexible implementation of topological codes but also enables the construction of topological codes with higher effective dimensions than those of the physical architecture. For example, in Fig.~\ref{fig:QEC}(b), a 1D cluster chain can be used to form a 2D [[12,2,3]] XZZX surface code~\cite{Roffe2023}, an error-correcting code specifically designed for biased noise. Notably, silicon-based spin qubit systems inherently exhibit such noise bias characteristics, with the qubit relaxation time significantly longer than the dephasing time~\cite{Pla2013,Muhonen2014,Tenberg2019,Thorvaldson2024, ZhangCH2025}. With such highly biased noise, the fidelity threshold for fault tolerance can be significantly lowered~\cite{ataides2021}. The practical performance of the XZZX code in the case of the donor-cluster-based system depends on the specific device. Further detailed QEC scheme and evaluation of error correction performance are subject to future study.

    \begin{figure*}[hbt!]
		\begin{center}
			\includegraphics[width=2\columnwidth]{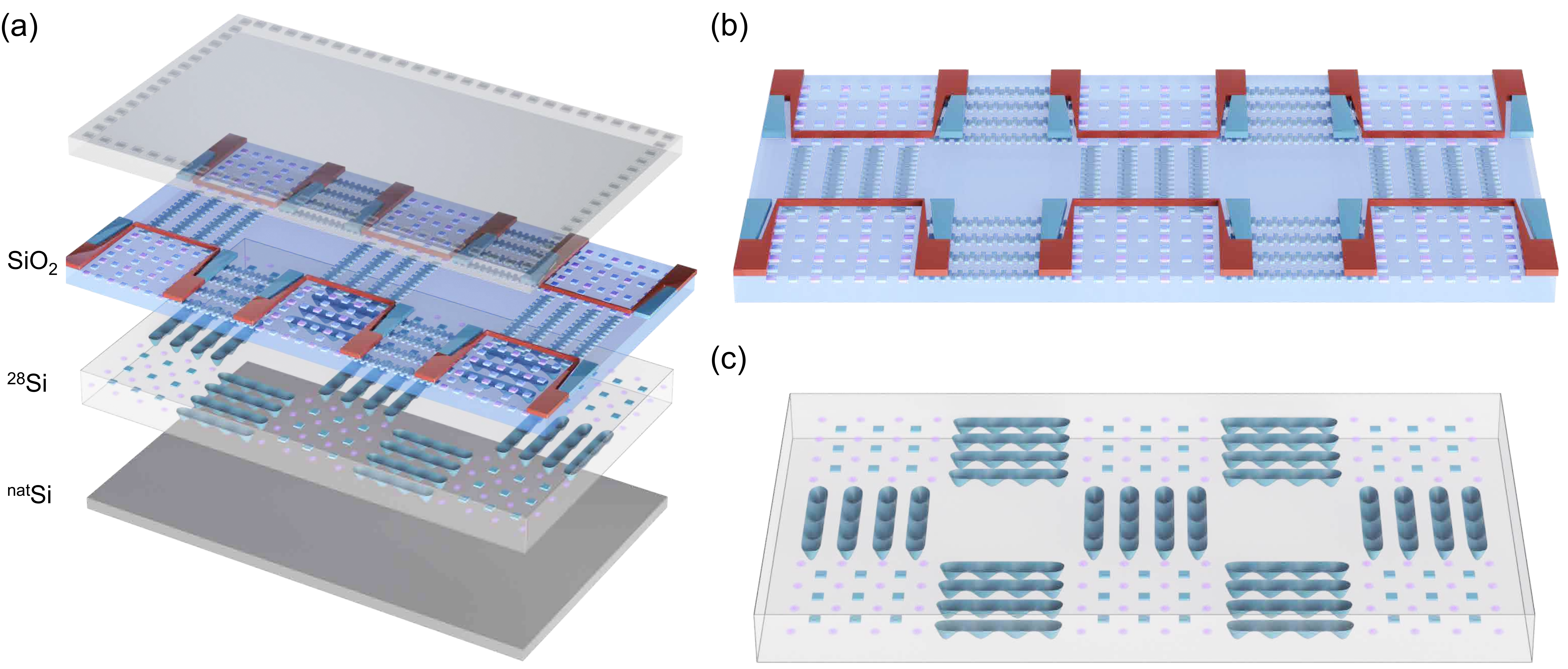}
		\end{center}
		\vspace{-0.50cm}
		\caption{A large-scale cluster-based quantum processor based on long-range couplings between clusters. (a) The large-scale quantum processor including multiple cluster arrays. The quantum processor consists of four layers: a classical electrical control layer at the top, a natural silicon substrate ($^{\rm{nat}}$Si) at the bottom, and two intermediate layers comprising a silicon dioxide (SiO$_{2}$) layer and a $^{28}$Si layer containing the cluster arrays. (b) Top view of the SiO$_{2}$ layer. This layer is the control layer, which contains control units for tuning and driving the underlying cluster array, including gates, micromagnets and antennas. One-dimensional chains formed by overlapping gates enables the control of moving QDs between the cluster arrays. (c) Top view of the $^{28}$Si layer. This layer includes the cluster arrays and the moving QDs, enabling the shuttling of electrons between arrays. The cluster arrays are interconnected by the shuttling-mediated coupling between electrons.}
		\label{fig:3D_moving_QD}
	\end{figure*}
    
    The connectivity between the qubits can further be enhanced by arranging the donor clusters in specific lattice configurations, such as the triangular lattice, and leveraging nearest-neighbor coupling. This enhanced connectivity provides high flexibility and opens up new possibilities for the design of QEC schemes. Furthermore, for 2D cluster arrays, QEC codes could also be extended to high-dimensional codes~\cite{Breuckmann2021,Quintavalle2021,HuangE2023}. Designing and optimizing QEC schemes specifically adapted to a 2D cluster array represents a valuable direction for further research. Considering the local all-to-all connectivity of qubits, the cluster architecture is also highly compatible with quantum low-density parity-check (qLDPC) codes, 
    which have demonstrated the capability to substantially reduce the physical resource overhead required for QEC~\cite{Roffe2023,Breuckmann2021,bravyi2024}. The qLDPC scheme based on a scalable cluster-based architecture represents another promising avenue for scalable FTQC.
    
    
    \subsection{Further scaling up of the system}
    The realization of FTQC encoding on the donor-cluster architecture requires further advancement of several critical technologies~\cite{shor1995,gottesman2010introduction}. First, enhanced tunability is needed for precise control of both HF interaction and exchange interaction, to enable high-fidelity gate operations. Fast and high-fidelity QND readout of nuclear spins within clusters must be realized, as it provides a critical foundation for mid-circuit measurement capabilities~\cite{kelly2015,krinner2022}, enabling real-time syndrome extraction and feedback control without disturbing the subsequent operations. Moreover, noise-mitigation techniques, such as dynamical decoupling and pulse optimization, should be integrated to suppress the error rate of qubit operations. These advancements will bridge the gap between the current cluster array implementations and a fully functional FTQC platform.

    To further scale the donor-cluster architecture, distributed cluster arrays can be interconnected, for example, via cQED cavities~\cite{Hu2012,Mi2018,borjans2020,Harvey-Collard2022,yu2023,dijkema2025} or electron shuttling techniques~\cite{taylor_fault-tolerant_2005,Huang2013,baart_single-spin_2016,zhao_doppler_2016,noiri_shuttling-based_2022,seidler2022,Langrock2023,xueR2024,de_smet2025}. Electron shuttling can be implemented via two schemes: one is bucket-brigade mode~\cite{taylor_fault-tolerant_2005,baart_single-spin_2016,seidler2022}, which transfers the electron along a 1D static QD chain by sequentially tuning electrochemical potential detuning and tunneling; the other is the conveyor mode~\cite{Huang2013,zhao_doppler_2016,noiri_shuttling-based_2022,Langrock2023,xueR2024,de_smet2025}, which transports the electron by applying voltage pulses to overlapping gates to form a moving QD in the silicon. For example, Figure \ref{fig:3D_moving_QD} shows a schematic diagram of a large-scale quantum processor formed by connecting cluster arrays through moving QDs. By utilizing interleaved gates to form moving QDs, coupling between nuclear spin qubits in distinct cluster arrays can be achieved through the shuttling of electron spin. In addition, the remaining space between the cluster arrays and the moving QDs can accommodate antennas, wiring fan-out, or classical control units. The interconnected cluster arrays provides a modular platform that allows for flexible design and connectivity. The flexibility enables the construction of various topologies required for FTQC.

    The cluster structure demonstrates unique advantages within phosphorus-donor nuclear spin systems. In the future, this architecture can also be adapted to other dopant species. For example, acceptor atoms such as boron or aluminum feature bound holes that permit fast EDSR control~\cite{Salfi2016PRL,ZhangSH2023}. Additionally, high-spin nuclei, such as antimony, arsenic, and aluminum, enable qudit~\cite{yuX2025} and logical qubit encoding~\cite{Gupta2024}. Extending the cluster paradigm to these systems is highly promising, as it could leverage their respective advantages to enhance processing capabilities for large-scale quantum computation in silicon. 

    \section{Conclusion}
	In conclusion, we propose a scalable spin qubit scheme based on donor cluster arrays, moving beyond the conventional single-dopant paradigm~\cite{Hill2015,Hill2021,ZhangSH2023,ZhangSH2025}. This scheme accommodates variations in the number of donors within clusters. Even if there are redundant nuclear spins that cannot be effectively utilized, their influence can be eliminated through initialization, thereby relaxing fabrication requirements. By utilizing electron spins as ancillas, addressable single-qubit operations on nuclear spin qubits can be achieved, along with intra- and inter-cluster multi-qubit gates. Incorporation of tunable exchange interactions, inhomogeneous HF distribution, and micromagnets provides a viable solution to the potential bottleneck of frequency crowding. Furthermore, through the strategic design of EA-TCMG and NA-TCMG, crosstalk errors can be effectively suppressed. Analytical and numerical evaluations demonstrate that both intra- and inter-cluster multi-qubit gate fidelities exceed 99\%. Our calculations quantify the impact of HF coupling, exchange interaction and magnetic field gradients on qubit gate fidelity, establishing experimental parameter requirements. To leverage the advantages of this scalable architecture, we propose viable QEC schemes tailored to the unique characteristics of cluster-based spin-qubit systems. 
    Our proposed donor-cluster array architecture offers a promising pathway toward realizing large-scale FTQC in silicon.

	\begin{acknowledgments}

    This work was supported by the National Natural Science Foundation of China (Grants No. 11904157 to P.H., 92165210 to Y.H., 62174076 to Y.H., and 12275117 to T.X.), Quantum Science and Technology-National Science and Technology Major Project (No. 2021ZD0302300 to Y.H., and 2024ZD0300400 to T.X.), Shenzhen Science and Technology Program (Grants No. KQTD20200820113010023 to Y.H.), and Guangdong Basic and Applied Basic Research Foundation (Grants No. 2022B1515020074 to T.X.).
\end{acknowledgments}

\bibliography{reff}

@PREAMBLE{
 "\providecommand{\noopsort}[1]{}" 
 # "\providecommand{\singleletter}[1]{#1}%" 
}

@article{tosato_crossbar_2026,
	title = {A crossbar chip for benchmarking semiconductor spin qubits},
	volume = {9},
	doi = {10.1038/s41928-026-01569-5},
	journal = {Nat. Electron.},
	author = {Tosato, Alberto and Elsayed, Asser and Poggiali, Federico and Stehouwer, Lucas Erik Adriaan and Costa, Davide and Hudson, Karina Louise and Degli Esposti, Davide and Scappucci, Giordano},
	year = {2026},
	pages = {324--333},
}

@article{ZhangCH2026,
	title = {Universal logical operations in a silicon quantum processor},
	issn = {1748-3395},
	doi = {10.1038/s41565-026-02140-1},
	journal = {Nat. Nanotechnol.},
	author = {Zhang, Chunhui and Xu, Feng and Zhang, Shihang and Duan, Mingchao and Zhong, Dupeng and Bai, Xuesong and Wang, Hao and Huang, Chao and Deng, Yi and Gao, Miao and Zhang, Yu-Ning and Liu, Jiaze and Li, Chunhui and Jiang, Yan and Zhao, Baolong and Shu, Huan and Wu, Kunrong and Shi, Keji and Ding, Qiming and Tian, Zhen and Wang, Guanyong and Yuan, Xiao and Xin, Tao and Hu, Guangchong and Liu, Song and Pan, Tianluo and Huang, Peihao and He, Yu and Yu, Dapeng},
	year = {2026},
}

@article{WangH2026,
	author={Wang, Hao and Zhao, Bao-Long and Zhang, Shi-Hang and Jin, Zi-Liang and Shu, Huan and Huang, Peihao and Wang, Guan-Yong and Duan, Ming-Chao and Pan, Tian-Luo and Xin, Tao and Hu, Guang-Chong and Tian, Zhen and Yu, Da-Peng and He, Yu},
	title={Micromagnet engineering for enhanced addressability of donor-cluster arrays in silicon},
	journal={Chin. Phys. B},
	doi={10.1088/1674-1056/ae4b29},
	year={2026},
}

@article{ZhangSH2025,
  title = {Addressable and tunable module for donor-based scalable silicon quantum computing},
  author = {Zhang, Shihang and He, Yu and Huang, Peihao},
  journal = {Phys. Rev. B},
  volume = {112},
  issue = {12},
  pages = {125303},
  numpages = {14},
  year = {2025},
  month = {Sep},
  publisher = {American Physical Society},
  doi = {10.1103/tfnf-b78h}
}

@article{yang_operation_2020,
	title = {Operation of a silicon quantum processor unit cell above one kelvin},
	volume = {580},
	issn = {1476-4687},
	url = {https://doi.org/10.1038/s41586-020-2171-6},
	doi = {10.1038/s41586-020-2171-6},
	number = {7803},
	journal = {Nature},
	author = {Yang, C. H. and Leon, R. C. C. and Hwang, J. C. C. and Saraiva, A. and Tanttu, T. and Huang, W. and Camirand Lemyre, J. and Chan, K. W. and Tan, K. Y. and Hudson, F. E. and Itoh, K. M. and Morello, A. and Pioro-Ladrière, M. and Laucht, A. and Dzurak, A. S.},
	month = apr,
	year = {2020},
	pages = {350--354},
}

@article{Tenberg2019,
  title = {Electron spin relaxation of single phosphorus donors in metal-oxide-semiconductor nanoscale devices},
  author = {Tenberg, Stefanie B. and Asaad, Serwan and M\k{a}dzik, Mateusz T. and Johnson, Mark A. I. and Joecker, Benjamin and Laucht, Arne and Hudson, Fay E. and Itoh, Kohei M. and Jakob, A. Malwin and Johnson, Brett C. and Jamieson, David N. and McCallum, Jeffrey C. and Dzurak, Andrew S. and Joynt, Robert and Morello, Andrea},
  journal = {Phys. Rev. B},
  volume = {99},
  issue = {20},
  pages = {205306},
  numpages = {13},
  year = {2019},
  month = {May},
  publisher = {American Physical Society},
  doi = {10.1103/PhysRevB.99.205306},
}

@article{noiri_shuttling-based_2022,
	title = {A shuttling-based two-qubit logic gate for linking distant silicon quantum processors},
	volume = {13},
	issn = {2041-1723},
	doi = {10.1038/s41467-022-33453-z},
	number = {1},
	journal = {Nat. Commun.},
	author = {Noiri, Akito and Takeda, Kenta and Nakajima, Takashi and Kobayashi, Takashi and Sammak, Amir and Scappucci, Giordano and Tarucha, Seigo},
	month = sep,
	year = {2022},
	pages = {5740},
}

@article{zhao_doppler_2016,
	title = {Doppler effect induced spin relaxation boom},
	volume = {6},
	issn = {2045-2322},
	doi = {10.1038/srep23169},
	number = {1},
	journal = {Sci. Rep.},
	author = {Zhao, Xinyu and Huang, Peihao and Hu, Xuedong},
	month = mar,
	year = {2016},
	pages = {23169},
}

@article{baart_single-spin_2016,
	title = {Single-spin {CCD}},
	volume = {11},
	issn = {1748-3395},
	doi = {10.1038/nnano.2015.291},
	number = {4},
	journal = {Nat. Nanotechnol.},
	author = {Baart, T. A. and Shafiei, M. and Fujita, T. and Reichl, C. and Wegscheider, W. and Vandersypen, L. M. K.},
	month = apr,
	year = {2016},
	pages = {330--334},
}

@article{taylor_fault-tolerant_2005,
	title = {Fault-tolerant architecture for quantum computation using electrically controlled semiconductor spins},
	volume = {1},
	issn = {1745-2481},
	doi = {10.1038/nphys174},
	number = {3},
	journal = {Nat. Phys.},
	author = {Taylor, J. M. and Engel, H.-A. and Dür, W. and Yacoby, A. and Marcus, C. M. and Zoller, P. and Lukin, M. D.},
	year = {2005},
	pages = {177--183},
}

@article{Mi2018,
	title = {A coherent spin–photon interface in silicon},
	volume = {555},
	issn = {1476-4687},
	doi = {10.1038/nature25769},
	number = {7698},
	journal = {Nature},
	author = {Mi, X. and Benito, M. and Putz, S. and Zajac, D. M. and Taylor, J. M. and Burkard, Guido and Petta, J. R.},
	month = mar,
	year = {2018},
	pages = {599--603},
}

@article{kranz2023use_of_exchange,
author = {Kranz, Ludwik and Gorman, Samuel K. and Thorgrimsson, Brandur and Monir, Serajum and He, Yu and Keith, Daniel and Charde, Keshavi and Keizer, Joris G. and Rahman, Rajib and Simmons, Michelle Y.},
title = {The use of exchange coupled atom qubits as atomic-scale magnetic field sensors},
journal = {Adv. Mater.},
volume = {35},
number = {6},
pages = {2201625},
doi = {https://doi.org/10.1002/adma.202201625},
year = {2023}
}

@article{ZhangCH2025,
	title = {Quantum error detection in a silicon quantum processor},
	volume = {9},
	doi = {10.1038/s41928-025-01557-1},
	journal = {Nat. Electron.},
	author = {Zhang, Chunhui and Li, Chunhui and Tian, Zhen and Jiang, Yan and Xu, Feng and Zhang, Shihang and Wang, Hao and Zhang, Yu-Ning and Bai, Xuesong and Zhao, Baolong and Zhang, Yi-Fei and Shu, Huan and Liu, Jiaze and Wu, Kunrong and Huang, Chao and Shi, Keji and Duan, Mingchao and Xin, Tao and Huang, Peihao and Pan, Tianluo and Liu, Song and Wang, Guanyong and Hu, Guangchong and He, Yu and Yu, Dapeng},
	year = {2026},
	pages = {295--303},
}

@article{Muhonen2015,
	title = {Quantifying the quantum gate fidelity of single-atom spin qubits in silicon by randomized benchmarking},
	volume = {27},
	doi = {10.1088/0953-8984/27/15/154205},
	number = {15},
	journal = {J. Phys. Condens. Matter},
	author = {Muhonen, J T and Laucht, A and Simmons, S and Dehollain, J P and Kalra, R and Hudson, F E and Freer, S and Itoh, K M and Jamieson, D N and McCallum, J C and Dzurak, A S and Morello, A},
	month = mar,
	year = {2015},
	pages = {154205},
}

@article{Kitaev1997universal_gate,
	author = {{Kitaev}, A. Yu},
	title = "{Quantum computations: algorithms and error correction}",
	journal = {Russ. Math. Surv.},
	year = 1997,
	volume = {52},
	number = {6},
	pages = {1191-1249},
	doi = {10.1070/RM1997v052n06ABEH002155},
}

@article{Barenco1995universal_gate,
	title = {Elementary gates for quantum computation},
	author = {Barenco, Adriano and Bennett, Charles H. and Cleve, Richard and DiVincenzo, David P. and Margolus, Norman and Shor, Peter and Sleator, Tycho and Smolin, John A. and Weinfurter, Harald},
	journal = {Phys. Rev. A},
	volume = {52},
	pages = {3457--3467},
	numpages = {0},
	year = {1995},
	publisher = {American Physical Society},
doi = {10.1103/PhysRevA.52.3457}
}

@misc{undseth2026weight,
  title={Weight-four parity checks with silicon spin qubits},
  author={Undseth, Brennan and Meggiato, Nicola and Wu, Yi-Hsien and Katiraee-Far, Sam R and Tryputen, Larysa and de Snoo, Sander L and Esposti, Davide Degli and Scappucci, Giordano and Greplov{\'a}, Eli{\v{s}}ka and Vandersypen, Lieven MK},
  howpublished={preprint at {https://arxiv.org/abs/2601.23267 (2026)}},
}

@misc{ademi2025distributing,
  title={Distributing entanglement between distant semiconductor qubit registers using a shared-control shuttling link},
  author={Ademi, Zarije and Bassi, Marion and Yu, C{\'e}cile X and Oosterhout, Stefan D and Matsumoto, Yuta and de Snoo, Sander L and Sammak, Amir and Vandersypen, Lieven MK and Scappucci, Giordano and D{\'e}prez, Corentin and others},
  howpublished={preprint at {https://arxiv.org/abs/2510.26860 (2025)}},
}

@misc{mcintyre2026theory,
  title={Theory of spin qubits and the path to scalability},
  author={McIntyre, ZM and Sarkar, Abhikbrata and Loss, Daniel},
  howpublished={preprint at {https://arxiv.org/abs/2604.13644 (2026)}},
}

@misc{dijkema2026simultaneous,
  title={Simultaneous operation of an 18-qubit modular array in germanium},
  author={Dijkema, Jurgen J and Zhang, Xin and Bardakas, Achilleas and Bouman, Daniel and Cuzzocrea, Alice and van Driel, David and Girardi, Davide and Stehouwer, Lucas EA and Scappucci, Giordano and Zwerver, Anne-Marije J and others},
  howpublished={preprint at {https://arxiv.org/abs/2604.01063 (2026)}},
}

@misc{wuYh2025,
	title={Simultaneous High-Fidelity Single-Qubit Gates in a Spin Qubit Array}, 
	author={Wu, Yi-Hsien and Camenzind, Leon C and B{\"u}tler, Patrick and Jin, Ik Kyeong and Noiri, Akito and Takeda, Kenta and Nakajima, Takashi and Kobayashi, Takashi and Scappucci, Giordano and Goan, Hsi-Sheng and others},
	howpublished = {Preprint at {https://arxiv.org/abs/2507.11918 (2025)}},
}

@article{donnelly2025,
  title = {Noise correlations in an atom-based quantum dot array},
  author = {Donnelly, M.B. and Rowlands, J. and Kranz, L. and Hsueh, Y.L. and Chung, Y. and Timofeev, A.V. and Geng, H. and Singh-Gregory, P. and Gorman, S.K. and Keizer, J.G. and Rahman, R. and Simmons, M.Y.},
  journal = {Phys. Rev. Appl.},
  volume = {23},
  issue = {6},
  pages = {064058},
  numpages = {9},
  year = {2025},
  publisher = {American Physical Society},
  doi = {10.1103/kr2l-c97c},
}

@article{Gupta2024,
  title = {Robust macroscopic Schr\"odinger's cat on a nucleus},
  author = {Gupta, Pragati and Vaartjes, Arjen and Yu, Xi and Morello, Andrea and Sanders, Barry C.},
  journal = {Phys. Rev. Res.},
  volume = {6},
  issue = {1},
  pages = {013101},
  numpages = {11},
  year = {2024},
  publisher = {American Physical Society},
  doi = {10.1103/PhysRevResearch.6.013101},
}

@article{yuX2025,
	title = {Schrödinger cat states of a nuclear spin qudit in silicon},
	volume = {21},
	issn = {1745-2481},
	doi = {10.1038/s41567-024-02745-0},
	number = {3},
	journal = {Nat. Phys.},
	author = {Yu, Xi and Wilhelm, Benjamin and Holmes, Danielle and Vaartjes, Arjen and Schwienbacher, Daniel and Nurizzo, Martin and Kringhøj, Anders and Blankenstein, Mark R. van and Jakob, Alexander M. and Gupta, Pragati and Hudson, Fay E. and Itoh, Kohei M. and Murray, Riley J. and Blume-Kohout, Robin and Ladd, Thaddeus D. and Anand, Namit and Dzurak, Andrew S. and Sanders, Barry C. and Jamieson, David N. and Morello, Andrea},
	year = {2025},
	pages = {362--367},
}

@article{ZhangSH2023,
  title = {Acceptor-based qubit in silicon with tunable strain},
  author = {Zhang, Shihang and He, Yu and Huang, Peihao},
  journal = {Phys. Rev. B},
  volume = {107},
  issue = {15},
  pages = {155301},
  numpages = {12},
  year = {2023},
  publisher = {American Physical Society},
  doi = {10.1103/PhysRevB.107.155301},
}

@article{borjans2020,
	title = {Resonant microwave-mediated interactions between distant electron spins},
	volume = {577},
	issn = {1476-4687},
	doi = {10.1038/s41586-019-1867-y},
	number = {7789},
	journal = {Nature},
	author = {Borjans, F. and Croot, X. G. and Mi, X. and Gullans, M. J. and Petta, J. R.},
	year = {2020},
	pages = {195--198},
}

@article{dijkema2025,
	title = {Cavity-mediated {iSWAP} oscillations between distant spins},
	volume = {21},
	issn = {1745-2481},
	doi = {10.1038/s41567-024-02694-8},
	number = {1},
	journal = {Nat. Phys.},
	author = {Dijkema, Jurgen and Xue, Xiao and Harvey-Collard, Patrick and Rimbach-Russ, Maximilian and de Snoo, Sander L. and Zheng, Guoji and Sammak, Amir and Scappucci, Giordano and Vandersypen, Lieven M. K.},
	year = {2025},
	pages = {168--174},
}

@article{yu2023,
	title = {Strong coupling between a photon and a hole spin in silicon},
	volume = {18},
	issn = {1748-3395},
	doi = {10.1038/s41565-023-01332-3},
	number = {7},
	journal = {Nat. Nanotechnol.},
	author = {Yu, Cecile X. and Zihlmann, Simon and Abadillo-Uriel, Jose C. and Michal, Vincent P. and Rambal, Nils and Niebojewski, Heimanu and Bedecarrats, Thomas and Vinet, Maud and Dumur, Etienne and Filippone, Michele and Bertrand, Benoit and De Franceschi, Silvano and Niquet, Yann-Michel and Maurand, Romain},
	year = {2023},
	pages = {741--746},
}

@article{Harvey-Collard2022,
  title = {Coherent Spin-Spin Coupling Mediated by Virtual Microwave Photons},
  author = {Harvey-Collard, Patrick and Dijkema, Jurgen and Zheng, Guoji and Sammak, Amir and Scappucci, Giordano and Vandersypen, Lieven M. K.},
  journal = {Phys. Rev. X},
  volume = {12},
  issue = {2},
  pages = {021026},
  numpages = {16},
  year = {2022},
  publisher = {American Physical Society},
  doi = {10.1103/PhysRevX.12.021026},
}

@article{xueR2024,
	title = {Si/{SiGe} {QuBus} for single electron information-processing devices with memory and micron-scale connectivity function},
	volume = {15},
	issn = {2041-1723},
	doi = {10.1038/s41467-024-46519-x},
	number = {1},
	journal = {Nat. Commun.},
	author = {Xue, Ran and Beer, Max and Seidler, Inga and Humpohl, Simon and Tu, Jhih-Sian and Trellenkamp, Stefan and Struck, Tom and Bluhm, Hendrik and Schreiber, Lars R.},
	year = {2024},
	pages = {2296},
}

@article{Langrock2023,
  title = {Blueprint of a Scalable Spin Qubit Shuttle Device for Coherent Mid-Range Qubit Transfer in Disordered $\text{Si/SiGe/SiO}_{2}$},
  author = {Langrock, Veit and Krzywda, Jan A. and Focke, Niels and Seidler, Inga and Schreiber, Lars R. and Cywi\ifmmode \acute{n}\else \'{n}\fi{}ski, \L{}ukasz},
  journal = {PRX Quantum},
  volume = {4},
  issue = {2},
  pages = {020305},
  numpages = {35},
  year = {2023},
  publisher = {American Physical Society},
  doi = {10.1103/PRXQuantum.4.020305},
}

@article{seidler2022,
	title = {Conveyor-mode single-electron shuttling in {Si}/{SiGe} for a scalable quantum computing architecture},
	volume = {8},
	issn = {2056-6387},
	doi = {10.1038/s41534-022-00615-2},
	number = {1},
	journal = {npj Quantum Inf.},
	author = {Seidler, Inga and Struck, Tom and Xue, Ran and Focke, Niels and Trellenkamp, Stefan and Bluhm, Hendrik and Schreiber, Lars R.},
	year = {2022},
	pages = {100},
}

@article{Huang2013,
  title = {Spin qubit relaxation in a moving quantum dot},
  author = {Huang, Peihao and Hu, Xuedong},
  journal = {Phys. Rev. B},
  volume = {88},
  issue = {7},
  pages = {075301},
  numpages = {9},
  year = {2013},
  publisher = {American Physical Society},
  doi = {10.1103/PhysRevB.88.075301},
}

@article{de_smet2025,
	title = {High-fidelity single-spin shuttling in silicon},
	volume = {20},
	issn = {1748-3395},
	doi = {10.1038/s41565-025-01920-5},
	number = {7},
	journal = {Nat. Nanotechnol.},
	author = {De Smet, Maxim and Matsumoto, Yuta and Zwerver, Anne-Marije J. and Tryputen, Larysa and de Snoo, Sander L. and Amitonov, Sergey V. and Katiraee-Far, Sam R. and Sammak, Amir and Samkharadze, Nodar and Gul, onder and Wasserman, Rick N. M. and Greplova, Eliska and Rimbach-Russ, Maximilian and Scappucci, Giordano and Vandersypen, Lieven M. K.},
	year = {2025},
	pages = {866--872},
}

@article{kelly2015,
	title = {State preservation by repetitive error detection in a superconducting quantum circuit},
	volume = {519},
	issn = {1476-4687},
	doi = {10.1038/nature14270},
	number = {7541},
	journal = {Nature},
	author = {Kelly, J. and Barends, R. and Fowler, A. G. and Megrant, A. and Jeffrey, E. and White, T. C. and Sank, D. and Mutus, J. Y. and Campbell, B. and Chen, Yu and Chen, Z. and Chiaro, B. and Dunsworth, A. and Hoi, I.-C. and Neill, C. and O’Malley, P. J. J. and Quintana, C. and Roushan, P. and Vainsencher, A. and Wenner, J. and Cleland, A. N. and Martinis, John M.},
	year = {2015},
	pages = {66--69},
}

@article{krinner2022,
	title = {Realizing repeated quantum error correction in a distance-three surface code},
	volume = {605},
	issn = {1476-4687},
	doi = {10.1038/s41586-022-04566-8},
	number = {7911},
	journal = {Nature},
	author = {Krinner, Sebastian and Lacroix, Nathan and Remm, Ants and Di Paolo, Agustin and Genois, Elie and Leroux, Catherine and Hellings, Christoph and Lazar, Stefania and Swiadek, Francois and Herrmann, Johannes and Norris, Graham J. and Andersen, Christian Kraglund and Muller, Markus and Blais, Alexandre and Eichler, Christopher and Wallraff, Andreas},
	year = {2022},
	pages = {669--674},
}

@article{HuangE2023,
  title = {Tailoring Three-Dimensional Topological Codes for Biased Noise},
  author = {Huang, Eric and Pesah, Arthur and Chubb, Christopher T. and Vasmer, Michael and Dua, Arpit},
  journal = {PRX Quantum},
  volume = {4},
  issue = {3},
  pages = {030338},
  numpages = {35},
  year = {2023},
  publisher = {American Physical Society},
  doi = {10.1103/PRXQuantum.4.030338},
}

@article{Quintavalle2021,
  title = {Single-Shot Error Correction of Three-Dimensional Homological Product Codes},
  author = {Quintavalle, Armanda O. and Vasmer, Michael and Roffe, Joschka and Campbell, Earl T.},
  journal = {PRX Quantum},
  volume = {2},
  issue = {2},
  pages = {020340},
  numpages = {32},
  year = {2021},
  publisher = {American Physical Society},
  doi = {10.1103/PRXQuantum.2.020340},
}

@article{wyrick2022,
	title = {Enhanced {Atomic} {Precision} {Fabrication} by {Adsorption} of {Phosphine} into {Engineered} {Dangling} {Bonds} on {H}–{Si} {Using} {STM} and {DFT}},
	volume = {16},
	number = {11},
	journal = {ACS Nano},
	author = {Wyrick, Jonathan and Wang, Xiqiao and Namboodiri, Pradeep and Kashid, Ranjit Vilas and Fei, Fan and Fox, Joseph and Silver, Richard},
	year = {2022},
	Publisher = {American Chemical Society},
	pages = {19114--19123},
    doi = {10.1021/acsnano.2c08162}
}

@article{ataides2021,
	title = {The {XZZX} surface code},
	volume = {12},
	issn = {2041-1723},
	url = {https://doi.org/10.1038/s41467-021-22274-1},
	doi = {10.1038/s41467-021-22274-1},
	number = {1},
	journal = {Nat. Commun.},
	author = {Bonilla Ataides, J. Pablo and Tuckett, David K. and Bartlett, Stephen D. and Flammia, Steven T. and Brown, Benjamin J.},
	year = {2021},
	pages = {2172},
}

@misc{gottesman2010introduction,
      title={\textnormal{An introduction to quantum error correction and fault-tolerant quantum computation.}}, 
      author={Daniel Gottesman},
      
      howpublished = {Preprint at {https://arxiv.org/abs/0904.2557 (2009)}},
}

@article{shor1995,
  title = {Scheme for reducing decoherence in quantum computer memory},
  author = {Shor, Peter W.},
  journal = {Phys. Rev. A},
  doi = {10.1103/PhysRevA.52.R2493},
  volume = {52},
  pages = {R2493--R2496},
  year = {1995}

}

@article{Breuckmann2021,
  title = {Quantum Low-Density Parity-Check Codes},
  author = {Breuckmann, Nikolas P. and Eberhardt, Jens Niklas},
  journal = {PRX Quantum},
  volume = {2},
  issue = {4},
  pages = {040101},
  numpages = {19},
  year = {2021},
  publisher = {American Physical Society},
  doi = {10.1103/PRXQuantum.2.040101},
}

@article{bravyi2024,
	title = {High-threshold and low-overhead fault-tolerant quantum memory},
	volume = {627},
	issn = {1476-4687},
	doi = {10.1038/s41586-024-07107-7},
	number = {8005},
	journal = {Nature},
	author = {Bravyi, Sergey and Cross, Andrew W. and Gambetta, Jay M. and Maslov, Dmitri and Rall, Patrick and Yoder, Theodore J.},
	year = {2024},
	pages = {778--782},
}

@article{Roffe2023,
  doi = {10.22331/q-2023-05-15-1005},
  title = {Bias-tailored quantum {LDPC} codes},
  author = {Roffe, Joschka and Cohen, Lawrence Z. and Quintavalle, Armanda O. and Chandra, Daryus and Campbell, Earl T.},
  journal = {{Quantum}},
  issn = {2521-327X},
  publisher = {{Verein zur F{\"{o}}rderung des Open Access Publizierens in den Quantenwissenschaften}},
  volume = {7},
  pages = {1005},
  year = {2023}
}

@article{Monir2024,
  title = {Impact of measurement backaction on nuclear spin qubits in silicon},
  author = {Monir, S. and Osika, E. N. and Gorman, S. K. and Thorvaldson, I. and Hsueh, Y.-L. and Macha, P. and Kranz, L. and Reiner, J. and Simmons, M. Y. and Rahman, R.},
  journal = {Phys. Rev. B},
  volume = {109},
  issue = {3},
  pages = {035157},
  numpages = {7},
  year = {2024},
  publisher = {American Physical Society},
  doi = {10.1103/PhysRevB.109.035157},
}

@article{ruffino2022,
	title = {A cryo-{CMOS} chip that integrates silicon quantum dots and multiplexed dispersive readout electronics},
	volume = {5},
	issn = {2520-1131},
	doi = {10.1038/s41928-021-00687-6},
	number = {1},
	journal = {Nat. Electron.},
	author = {Ruffino, Andrea and Yang, Tsung-Yeh and Michniewicz, John and Peng, Yatao and Charbon, Edoardo and Gonzalez-Zalba, Miguel Fernando},
	year = {2022},
	pages = {53--59},
}

@article{steinacker2025,
	title = {Industry-compatible silicon spin-qubit unit cells exceeding 99\% fidelity},
	issn = {1476-4687},
    volume = {646},
    pages = {81--87},
	journal = {Nature},
    doi = {10.1038/s41586-025-09531-9},
	author = {Steinacker, Paul and Dumoulin Stuyck, Nard and Lim, Wee Han and Tanttu, Tuomo and Feng, MengKe and Serrano, Santiago and Nickl, Andreas and Candido, Marco and Cifuentes, Jesus D. and Vahapoglu, Ensar and Bartee, Samuel K. and Hudson, Fay E. and Chan, Kok Wai and Kubicek, Stefan and Jussot, Julien and Canvel, Yann and Beyne, Sofie and Shimura, Yosuke and Loo, Roger and Godfrin, Clement and Raes, Bart and Baudot, Sylvain and Wan, Danny and Laucht, Arne and Yang, Chih Hwan and Saraiva, Andre and Escott, Christopher C. and De Greve, Kristiaan and Dzurak, Andrew S.},
	year = {2025},
    }

@article{edlbauer2025,
	title = {An 11-qubit atom processor in silicon},
	volume = {648},
	issn = {1476-4687},
	doi = {10.1038/s41586-025-09827-w},
	journal = {Nature},
	author = {Edlbauer, Hermann and Wang, Junliang and Huq, A. M. Saffat-Ee and Thorvaldson, Ian and Jones, Michael T. and Misha, Saiful Haque and Pappas, William J. and Moehle, Christian M. and Hsueh, Yu-Ling and Bornemann, Henric and Gorman, Samuel K. and Chung, Yousun and Keizer, Joris G. and Kranz, Ludwik and Simmons, Michelle Y.},
	year = {2025},
	pages = {569--575},
}

@article{stemp2025,
	title = {Scalable entanglement of nuclear spins mediated by electron exchange},
	volume = {389},
	doi = {10.1126/science.ady3799},
	number = {6766},
	journal = {Science},
	author = {Stemp, Holly G. and van Blankenstein, Mark R. and Asaad, Serwan and Madzik, Mateusz T. and Joecker, Benjamin and Firgau, Hannes R. and Laucht, Arne and Hudson, Fay E. and Dzurak, Andrew S. and Itoh, Kohei M. and Jakob, Alexander M. and Johnson, Brett C. and Jamieson, David N. and Morello, Andrea},
	month = sep,
	year = {2025},
	Publisher ={ American Association for the Advancement of Science},
	pages = {1234--1238},
}

@article{Kane1998,
  title={A silicon-based nuclear spin quantum computer},
  author={Kane, Bruce E},
  journal={Nature},
  volume={393},
  number={6681},
  pages={133--137},
  year={1998},
  DOI={10.1038/30156},
  publisher={Nature Publishing Group}
}

@article{Golovach2006,
  title = {Electric-dipole-induced spin resonance in quantum dots},
  author = {Golovach, Vitaly N. and Borhani, Massoud and Loss, Daniel},
  journal = {Phys. Rev. B},
  volume = {74},
  issue = {16},
  pages = {165319},
  numpages = {10},
  year = {2006},
  month = {Oct},
  publisher = {American Physical Society},
  doi = {10.1103/PhysRevB.74.165319},
  url = {https://link.aps.org/doi/10.1103/PhysRevB.74.165319}
}

@article{Hu2012,
	author = {Hu, Xuedong and Liu, Yu-xi and Nori, Franco},
	title = {Strong coupling of a spin qubit to a superconducting stripline cavity},
	journal = {Phys. Rev. B},
	volume = {86},
	number = {3},
    pages = {035314},
	ISSN = {1098-0121
	1550-235X},
	DOI = {10.1103/PhysRevB.86.035314},
	year = {2012},
	type = {Journal Article}
}

@article{Loss1998,
  title = {Quantum computation with quantum dots},
  author = {Loss, Daniel and DiVincenzo, David P.},
  journal = {Phys. Rev. A},
  volume = {57},
  issue = {1},
  pages = {120--126},
  numpages = {0},
  year = {1998},
  month = {Jan},
  publisher = {American Physical Society},
  doi = {10.1103/PhysRevA.57.120},
  url = {https://link.aps.org/doi/10.1103/PhysRevA.57.120}
}

@article{Salfi2016PRL,
  title = {Charge-Insensitive Single-Atom Spin-Orbit Qubit in Silicon},
  author = {Salfi, Joe and Mol, Jan A. and Culcer, Dimitrie and Rogge, Sven},
  journal = {Phys. Rev. Lett.},
  volume = {116},
  issue = {24},
  pages = {246801},
  numpages = {6},
  year = {2016},
  month = {Jun},
  publisher = {American Physical Society},
  doi = {10.1103/PhysRevLett.116.246801},
  url = {https://link.aps.org/doi/10.1103/PhysRevLett.116.246801}
}

@article{Zwanenburg2013,
  title = {Silicon quantum electronics},
  author = {Zwanenburg, Floris A. and Dzurak, Andrew S. and Morello, Andrea and Simmons, Michelle Y. and Hollenberg, Lloyd C. L. and Klimeck, Gerhard and Rogge, Sven and Coppersmith, Susan N. and Eriksson, Mark A.},
  journal = {Rev. Mod. Phys.},
  volume = {85},
  issue = {3},
  pages = {961--1019},
  numpages = {0},
  year = {2013},
  month = {Jul},
  publisher = {American Physical Society},
  doi = {10.1103/RevModPhys.85.961},
  url = {https://link.aps.org/doi/10.1103/RevModPhys.85.961}
}

@article{Burkard2021,
  title = {Semiconductor spin qubits},
  author = {Burkard, Guido and Ladd, Thaddeus D. and Pan, Andrew and Nichol, John M. and Petta, Jason R.},
  journal = {Rev. Mod. Phys.},
  volume = {95},
  issue = {2},
  pages = {025003},
  numpages = {58},
  year = {2023},
  month = {Jun},
  publisher = {American Physical Society},
  doi = {10.1103/RevModPhys.95.025003},
}

@article{tyryshkin2012,
  title={Electron spin coherence exceeding seconds in high-purity silicon},
  author={Tyryshkin, Alexei M and Tojo, Shinichi and Morton, John JL and Riemann, Helge and Abrosimov, Nikolai V and Becker, Peter and Pohl, Hans-Joachim and Schenkel, Thomas and Thewalt, Michael LW and Itoh, Kohei M and others},
  journal={Nat. Mater.},
  volume={11},
  number={2},
  pages={143--147},
  year={2012},
  doi={10.1038/nmat3182},
  publisher={Nature Publishing Group},
}

@article{Tokura2006,
  title = {Coherent Single Electron Spin Control in a Slanting {Zeeman} Field},
  author = {Tokura, Yasuhiro and van der Wiel, Wilfred G. and Obata, Toshiaki and Tarucha, Seigo},
  journal = {Phys. Rev. Lett.},
  volume = {96},
  issue = {4},
  pages = {047202},
  numpages = {4},
  year = {2006},
  month = {Jan},
  publisher = {American Physical Society},
  doi = {10.1103/PhysRevLett.96.047202},
  url = {https://link.aps.org/doi/10.1103/PhysRevLett.96.047202}
}

@article{Tosi2017,
   author = {Tosi, G. and Mohiyaddin, F. A. and Schmitt, V. and Tenberg, S. and Rahman, R. and Klimeck, G. and Morello, A.},
   title = {Silicon quantum processor with robust long-distance qubit couplings},
   journal = {Nat. Commun.},
   volume = {8},
   number = {1},
   pages = {450},
   ISSN = {2041-1723 (Electronic)
2041-1723 (Linking)},
   DOI = {10.1038/s41467-017-00378-x},
   url = {https://www.ncbi.nlm.nih.gov/pubmed/28878207},
   year = {2017},
   type = {Journal Article}
}

@article{Madzik2022,
   author = {M{\c a}dzik, Mateusz T. and Asaad, Serwan and Youssry, Akram and Joecker, Benjamin and Rudinger, Kenneth M. and Nielsen, Erik and Young, Kevin C. and Proctor, Timothy J. and Baczewski, Andrew D. and Laucht, Arne and Schmitt, Vivien and Hudson, Fay E. and Itoh, Kohei M. and Jakob, Alexander M. and Johnson, Brett C. and Jamieson, David N. and Dzurak, Andrew S. and Ferrie, Christopher and Blume-Kohout, Robin and Morello, Andrea},
   title = {Precision tomography of a three-qubit donor quantum processor in silicon},
   journal = {Nature},
   volume = {601},
   number = {7893},
   pages = {348-353},
   ISSN = {1476-4687},
   DOI = {10.1038/s41586-021-04292-7},
   year = {2022},
   type = {Journal Article}
}

@article{Pla2013,
  title={High-fidelity readout and control of a nuclear spin qubit in silicon},
  author={Pla, Jarryd J and Tan, Kuan Y and Dehollain, Juan P and Lim, Wee H and Morton, John JL and Zwanenburg, Floris A and Jamieson, David N and Dzurak, Andrew S and Morello, Andrea},
  journal={Nature},
  volume={496},
  number={7445},
  pages={334--338},
  year={2013},
  publisher={Nature Publishing Group},
  doi={10.1038/nature12011},
}

@misc{supp,
	title = {See the {Supplemental Materials} at [url will be inserted
	by publisher], which provides discussions of architecture implmentation utilizing tuanble couplers or state-of-the-art techniques, the optimal number of donors within a cluster, and tuning-up of the system. {It} also introduces the analytical and numerical methods used in this work, presents a detailed analysis of the performance of the {ESR}-based {CZ} gate, and further discusses {ESR}- and {NMR}-assisted inter-cluster {CZ} gate schemes. {The Supplemental Material} also contains {Refs.} [79-83]}
}

@article{Muhonen2014,
   author = {Muhonen, Juha T. and Dehollain, Juan P. and Laucht, Arne and Hudson, Fay E. and Kalra, Rachpon and Sekiguchi, Takeharu and Itoh, Kohei M. and Jamieson, David N. and McCallum, Jeffrey C. and Dzurak, Andrew S. and Morello, Andrea},
   title = {Storing quantum information for 30 seconds in a nanoelectronic device},
   journal = {Nat. Nanotechnol.},
   volume = {9},
   number = {12},
   pages = {986-991},
   ISSN = {1748-3395},
   DOI = {10.1038/nnano.2014.211},
   year = {2014},
   type = {Journal Article}
}

@article{
	Hill2015,
	author = {Charles D. Hill  and Eldad Peretz  and Samuel J. Hile  and Matthew G. House  and Martin Fuechsle  and Sven Rogge  and Michelle Y. Simmons  and Lloyd C. L. Hollenberg },
	title = {A surface code quantum computer in silicon},
	journal = {Sci. Adv.},
	volume = {1},
	number = {9},
	pages = {e1500707},
	year = {2015},
	doi = {10.1126/sciadv.1500707},
	URL = {https://www.science.org/doi/abs/10.1126/sciadv.1500707}}

@article{Hill2021,
	title={An exchange-based surface-code quantum computer architecture in silicon},
	author = {{Hill}, Charles D. and {Usman}, Muhammad and {Hollenberg}, Lloyd C.~L.},
	journal={arXiv preprint arXiv:2107.11981},
	year={2021},
	url = {
	https://doi.org/10.48550/arXiv.2107.11981
	},
}

@article{Baart2017,
	author = {Baart, T. A. and Fujita, T. and Reichl, C. and Wegscheider, W. and Vandersypen, L. M.},
	title = {Coherent spin-exchange via a quantum mediator},
	journal = {Nat. Nanotechnol.},
	volume = {12},
	number = {1},
	pages = {26-30},
	ISSN = {1748-3395 (Electronic)
	1748-3387 (Linking)},
	DOI = {10.1038/nnano.2016.188},
	url = {https://www.ncbi.nlm.nih.gov/pubmed/27723732},
	year = {2017},
	type = {Journal Article}
}

@article{Buch2013,
	author = {Buch, H. and Mahapatra, S. and Rahman, R. and Morello, A. and Simmons, M. Y.},
	title = {Spin readout and addressability of phosphorus-donor clusters in silicon},
	journal = {Nat. Commun.},
	volume = {4},
	pages = {2017},
	ISSN = {2041-1723 (Electronic)
	2041-1723 (Linking)},
	DOI = {10.1038/ncomms3017},
	url = {https://www.ncbi.nlm.nih.gov/pubmed/23774081},
	year = {2013},
	type = {Journal Article}
}

@article{Kranz2023,
	author = {Kranz, Ludwik and Roche, Stephen and Gorman, Samuel K. and Keizer, Joris G. and Simmons, Michelle Y.},
	title = {High-Fidelity {CNOT} Gate for Donor Electron Spin Qubits in Silicon},
	journal = {Phys. Rev. Appl.},
	volume = {19},
	number = {2},
	pages = {024068},
	ISSN = {2331-7019},
	DOI = {10.1103/PhysRevApplied.19.024068},
	url = {https://link.aps.org/doi/10.1103/PhysRevApplied.19.024068},
	year = {2023},
	type = {Journal Article}
}

@article{Malinowski2018,
  title = {Spin of a Multielectron Quantum Dot and Its Interaction with a Neighboring Electron},
  author = {Malinowski, Filip K. and Martins, Frederico and Smith, Thomas B. and Bartlett, Stephen D. and Doherty, Andrew C. and Nissen, Peter D. and Fallahi, Saeed and Gardner, Geoffrey C. and Manfra, Michael J. and Marcus, Charles M. and Kuemmeth, Ferdinand},
  journal = {Phys. Rev. X},
  volume = {8},
  issue = {1},
  pages = {011045},
  numpages = {23},
  year = {2018},
  month = {Mar},
  publisher = {American Physical Society},
  doi = {10.1103/PhysRevX.8.011045},
  url = {https://link.aps.org/doi/10.1103/PhysRevX.8.011045}
}

@article{Srinivasa2015,
	author = {Srinivasa, V. and Xu, H. and Taylor, J. M.},
	title = {Tunable Spin-Qubit Coupling Mediated by a Multielectron Quantum Dot},
	journal = {Phys. Rev. Lett.},
	volume = {114},
	number = {22},
	pages = {226803},
	ISSN = {1079-7114 (Electronic)
	0031-9007 (Linking)},
	DOI = {10.1103/PhysRevLett.114.226803},
	url = {https://www.ncbi.nlm.nih.gov/pubmed/26196638},
	year = {2015},
	type = {Journal Article}
}

@article{Munia2023,
  title = {Superexchange coupling of donor qubits in silicon},
  author = {Munia, Mushita M. and Monir, Serajum and Osika, Edyta N. and Simmons, Michelle Y. and Rahman, Rajib},
  journal = {Phys. Rev. Appl.},
  volume = {21},
  issue = {1},
  pages = {014038},
  numpages = {12},
  year = {2024},
  month = {Jan},
  publisher = {American Physical Society},
  doi = {10.1103/PhysRevApplied.21.014038},
  url = {https://link.aps.org/doi/10.1103/PhysRevApplied.21.014038}
}

@article{Fuechsle2012,
   author = {Fuechsle, Martin and Miwa, Jill A. and Mahapatra, Suddhasatta and Ryu, Hoon and Lee, Sunhee and Warschkow, Oliver and Hollenberg, Lloyd C. L. and Klimeck, Gerhard and Simmons, Michelle Y.},
   title = {A single-atom transistor},
   journal = {Nat. Nanotechnol.},
   volume = {7},
   number = {4},
   pages = {242-246},
   ISSN = {1748-3395},
   DOI = {10.1038/nnano.2012.21},
   url = {https://doi.org/10.1038/nnano.2012.21},
   year = {2012},
   type = {Journal Article}
}

@article{Takeda2021,
   author = {Takeda, Kenta and Noiri, Akito and Nakajima, Takashi and Yoneda, Jun and Kobayashi, Takashi and Tarucha, Seigo},
   title = {Quantum tomography of an entangled three-qubit state in silicon},
   journal = {Nat. Nanotechnol.},
   volume = {16},
   number = {9},
   pages = {965-969},
   ISSN = {1748-3395},
   DOI = {10.1038/s41565-021-00925-0},
   url = {https://doi.org/10.1038/s41565-021-00925-0},
   year = {2021},
   type = {Journal Article}
}

@article{
WangCA2024,
author = {Chien-An Wang  and Valentin John  and Hanifa Tidjani  and Cécile X. Yu  and Alexander S. Ivlev  and Corentin Déprez  and Floor van Riggelen-Doelman  and Benjamin D. Woods  and Nico W. Hendrickx  and William I. L. Lawrie  and Lucas E. A. Stehouwer  and Stefan D. Oosterhout  and Amir Sammak  and Mark Friesen  and Giordano Scappucci  and Sander L. de Snoo  and Maximilian Rimbach-Russ  and Francesco Borsoi  and Menno Veldhorst },
title = {Operating semiconductor quantum processors with hopping spins},
journal = {Science},
volume = {385},
number = {6707},
pages = {447-452},
year = {2024},
doi = {10.1126/science.ado5915},
URL = {https://www.science.org/doi/abs/10.1126/science.ado5915}}

@article{
Saeedi2013,
author = {Kamyar Saeedi  and Stephanie Simmons  and Jeff Z. Salvail  and Phillip Dluhy  and Helge Riemann  and Nikolai V. Abrosimov  and Peter Becker  and Hans-Joachim Pohl  and John J. L. Morton  and Mike L. W. Thewalt },
title = {Room-Temperature Quantum Bit Storage Exceeding 39 Minutes Using Ionized Donors in Silicon-28},
journal = {Science},
volume = {342},
number = {6160},
pages = {830-833},
year = {2013},
doi = {10.1126/science.1239584},
URL = {https://www.science.org/doi/abs/10.1126/science.1239584}}

@article{HuangJY2024,
   author = {Huang, Jonathan Y. and Su, Rocky Y. and Lim, Wee Han and Feng, MengKe and van Straaten, Barnaby and Severin, Brandon and Gilbert, Will and Dumoulin Stuyck, Nard and Tanttu, Tuomo and Serrano, Santiago and Cifuentes, Jesus D. and Hansen, Ingvild and Seedhouse, Amanda E. and Vahapoglu, Ensar and Leon, Ross C. C. and Abrosimov, Nikolay V. and Pohl, Hans-Joachim and Thewalt, Michael L. W. and Hudson, Fay E. and Escott, Christopher C. and Ares, Natalia and Bartlett, Stephen D. and Morello, Andrea and Saraiva, Andre and Laucht, Arne and Dzurak, Andrew S. and Yang, Chih Hwan},
   title = {{High-fidelity spin qubit operation and algorithmic initialization above 1 K}},
   journal = {Nature},
   volume = {627},
   number = {8005},
   pages = {772-777},
   ISSN = {1476-4687},
   DOI = {10.1038/s41586-024-07160-2},
   url = {https://doi.org/10.1038/s41586-024-07160-2},
   year = {2024},
   type = {Journal Article}
}

@article{Philips2022,
   author = {Philips, Stephan G. J. and Mądzik, Mateusz T. and Amitonov, Sergey V. and de Snoo, Sander L. and Russ, Maximilian and Kalhor, Nima and Volk, Christian and Lawrie, William I. L. and Brousse, Delphine and Tryputen, Larysa and Wuetz, Brian Paquelet and Sammak, Amir and Veldhorst, Menno and Scappucci, Giordano and Vandersypen, Lieven M. K.},
   title = {Universal control of a six-qubit quantum processor in silicon},
   journal = {Nature},
   volume = {609},
   number = {7929},
   pages = {919-924},
   ISSN = {1476-4687},
   DOI = {10.1038/s41586-022-05117-x},
   url = {https://doi.org/10.1038/s41586-022-05117-x},
   year = {2022},
   type = {Journal Article}
}

@article{Thorvaldson2024,
	title = {Grover’s algorithm in a four-qubit silicon processor above the fault-tolerant threshold},
	volume = {20},
	issn = {1748-3395},
	url = {https://doi.org/10.1038/s41565-024-01853-5},
	doi = {10.1038/s41565-024-01853-5},
	number = {4},
	journal = {Nat. Nanotechnol.},
	author = {Thorvaldson, I. and Poulos, D. and Moehle, C. M. and Misha, S. H. and Edlbauer, H. and Reiner, J. and Geng, H. and Voisin, B. and Jones, M. T. and Donnelly, M. B. and Peña, L. F. and Hill, C. D. and Myers, C. R. and Keizer, J. G. and Chung, Y. and Gorman, S. K. and Kranz, L. and Simmons, M. Y.},
	month = apr,
	year = {2025},
	pages = {472--477},
}

@article{hogg2023single,
  title={Single-shot readout of multiple donor electron spins with a gate-based sensor},
  author={Hogg, Mark R and Pakkiam, Prasanna and Gorman, Samuel K and Timofeev, Andrey V and Chung, Yousun and Gulati, Gurpreet K and House, Matthew G and Simmons, Michelle Y},
  journal={PRX Quantum},
  volume={4},
  number={1},
  pages={010319},
  year={2023},
  publisher={APS},
    doi = {10.1103/PRXQuantum.4.010319},
}

@article{volk2019loading,
  title={Loading a quantum-dot based “{Qubyte}” register},
  author={Volk, Christian and Zwerver, Anne-Marije J and Mukhopadhyay, Uditendu and Eendebak, Pieter T and van Diepen, Cornelis Jacobus and Dehollain, Juan Pablo and Hensgens, Toivo and Fujita, Takafumi and Reichl, Christian and Wegscheider, Werner and others},
  journal={npj Quantum Inf.},
  volume={5},
  number={1},
  pages={29},
  year={2019},
  publisher={Nature Publishing Group UK London},
doi = {10.1038/s41534-019-0146-y}
}

@article{mills2019shuttling,
  title={Shuttling a single charge across a one-dimensional array of silicon quantum dots},
  author={Mills, AR and Zajac, DM and Gullans, MJ and Schupp, FJ and Hazard, TM and Petta, Jason R},
  journal={Nat. commun.},
  volume={10},
  number={1},
  pages={1063},
  year={2019},
  publisher={Nature Publishing Group UK London},
doi = {10.1038/s41467-019-08970-z}
}

@article{krzywda2025qdarts,
  title={QDarts: A quantum dot array transition simulator for finding charge transitions in the presence of finite tunnel couplings, non-constant charging energies and sensor dots},
  author={Krzywda, Jan and Liu, Weikun and van Nieuwenburg, Evert and Krause, Oswin},
  journal={SciPost Phys. Codebases},
  pages={43},
  year={2025},
doi = {10.21468/SciPostPhysCodeb.43}
}



\clearpage
\begin{widetext}
	\appendix
	\setcounter{figure}{0}
	\setcounter{equation}{0}
    \renewcommand{\theHfigure}{S\arabic{figure}}
\renewcommand{\thefigure}{S\arabic{figure}}

    \renewcommand{\theHequation}{\arabic{equation}}
	\section*{Supplementary Materials}
    \section{Architecture implementation and system calibration}
    \subsection{Scalable cluster-based qubit scheme with tunable couplers}\label{sec:QD}
    \begin{figure}[hbt!]
	   \begin{center}
			\includegraphics[width=0.75\columnwidth]{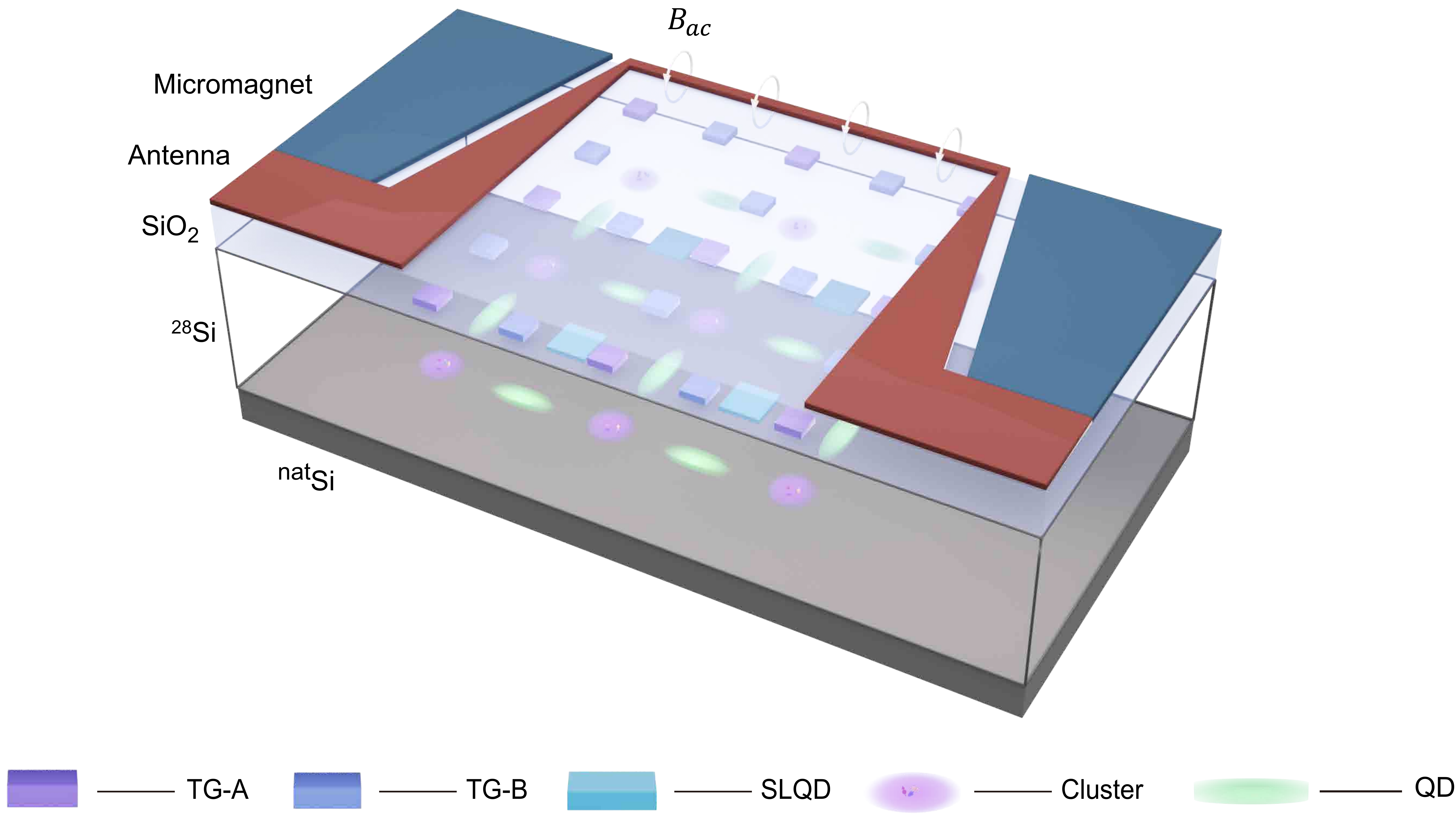}
		\end{center}
		\vspace{-0.50cm}
		\caption{A quantum processor based on the scalable cluster qubit scheme with tunable superexchange coupling between adjacent electrons. The cluster array is positioned in the  $^{28}$Si layer on top of the natural silicon ($^{\rm{nat}}$Si) layer. Each cluster contains a varying number of nuclei (represented by colored arrows) and a shared bound electron (depicted as `cloud’ surrounding the nuclei). The coupling between adjacent electron spins is mediated by the QDs (green) between clusters. The single-lead quantum dots (SLQDs) between clusters, represented as light blue box, is used for electron loading and readout. The properties of SLQDs are regulated by the control lines above them. Other elements within the cluster layer are controlled by the top gates (TGs) embedded in the upper oxide layer, with signals input through their connected control lines. At the top of the substrate are the micromagnets that provide the magnetic field gradient and the antennas that generate the alternating magnetic field $B_{ac}$.}
		\label{fig:3D_QD}
	\end{figure}
    To prevent crosstalk induced by frequency crowding, it is essential to turn off the effective exchange interaction between electron spins. According to the results shown in Sec.~\ref{sec: gate_performance} in the main text, the on-off ratio of the exchange interaction should be at least on the order of 10. Conventionally, the exchange interaction is tuned by adjusting the energy detuning between two adjacent sites. However, this approach offers limited tunability. Moreover, when the detuning is adjusted away from the charge symmetry point, the exchange interaction becomes susceptible to charge noise, resulting in errors. Therefore, alternative approaches are needed to overcome these limitations. There are two alternative approaches: one utilizes top gates (TGs) as barrier gates to adjust the tunneling between clusters, while the other using additional quantum dots (QDs) to mediate the highly tunable superexchange interactions between electron spins~\cite{Srinivasa2015,Malinowski2018,Baart2017,Munia2023,ZhangSH2025}. Fig.~\ref{fig:3D_QD} illustrates the architecture of a quantum processor based on the cluster qubit scheme with tunable superexchange couplings. The QDs between clusters serve not only as tunable couplers but can also be coupled to single-lead quantum dots (SLQDs) to enable indirect readout of electron spins on the clusters. Such quantum non-demolition readout of electron spin facilitates high-fidelity qubit readout.
    
	\subsection{Scalable cluster-based qubit scheme under state-of-the-art techniques}
	Indeed, to realize a fully scalable 2D architecture, several experimental techniques still require verification and breakthroughs, such as top-gate control, mid-circuit readout, and narrow gates. Nevertheless, the implementation of 1D cluster chains remains feasible. Moreover, owing to the local all-to-all connectivity within cluster systems, even 1D chains can demonstrate error-correction codes like the XZZX surface code. Fig.~\ref{fig:device} illustrates the spin-qubit device based on a 1D cluster chain achievable with state-of-the-art technologies. In the architecture,  in-plane gates are distributed on opposite sides of the chain to minimize crosstalk, and are utilized to adjust the electrochemical potentials of both the clusters and SLQDs. Readout and initialization of qubits can be achieved with mediation of electrons and SLQDs. Additionally, a micromagnet is incorporated to generate the required magnetic field gradient. The 1D cluster architecture not only provides sufficient qubits for near-term demonstration applications but also establishes the technical foundation and feasibility proof for future 2D scalable implementations.

    In the state-of-the-art device, the exchange interaction is tuned by adjusting the energy detuning between clusters. In this way, when the exchange interaction is turned on, the effect of charge noise on the exchange interaction is enhanced. Consequently, the fidelities of the inter-cluster qubit operations are reduced. We calculate the infidelities of the intra-cluster CNOT gate and inter-cluster Toffoli gate in a 2P-3P cluster system, same with the main text. Here, the noise-induced exchange interaction variation is considered. The tunneling rate between clusters is set to 3.6 GHz. The Coulomb potential $U$ is 30 meV. The corresponding exchange coupling at the charge symmetry point is approximately 7 MHz. As shown in Fig.~\ref{fig:device}(b), the infidelity of the intra-cluster CNOT gate is similar with that without considering exchange variation (Fig.~\ref{fig:gate_ct}(b) in the main text). This is because the sensitivity of the exchange interaction on charge noise is suppressed around the charge symmetry point. However, as shown in Fig.~\ref{fig:device}(c), the fidelity of the inter-cluster Toffoli gate is significantly reduced compared to Fig.~\ref{fig:gate_ct}(e) in the main text. Moreover, the selectable range of both the exchange interaction coupling and the HF difference is reduced, which imposes more stringent requirements on the device fabrication. The using of the TG or superexchange plays an important role for the scaling-up of the cluster-based system.
	\begin{figure}[htbp]
		\begin{center}
			\includegraphics[width=0.9\columnwidth]{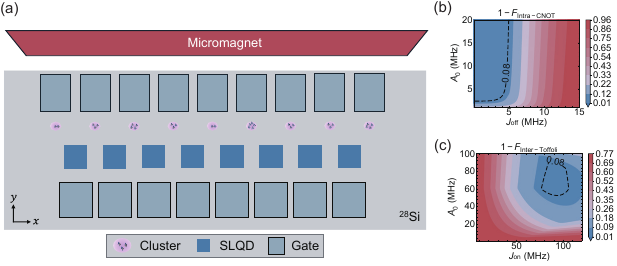}
		\end{center}
		\vspace{-0.50cm}
		\caption{Scalable cluster-based spin qubit scheme in $^{28}$Si. Donor clusters are arranged into a one-dimensional chain. Each donor cluster contains multiple P donors and one electron bound to them. To achieve tunability of exchange interactions between neighboring electrons, the inter-cluster units can be implemented either as top gates (TGs) for adjusting tunneling or as quantum dot (QDs) to induce tunable superexchange interactions. Additionally, single-lead quantum dots (SLQDs) are incorporated between clusters for readout and loading of electrons. The control gates for the clusters and the SLQDs are arranged on opposite sides (upper and lower) of the cluster chain. Moreover, a micromagnet is positioned adjacent to the cluster chain to generate magnetic field gradients, which mitigate the frequency crowding issue. (b) The infidelity $1-F_{\rm{Intra-CNOT}}$ of the intra-cluster CNOT-type gate is plotted as a function of the residual exchange interaction $J_{\rm{off}}$ and the hyperfine interaction $A_{0}$. (c) The infidelity $1-F_{\rm{Inter-Toffoli}}$ of the inter-cluster Toffoli-type gate is plotted as a function of the activated exchange interaction $J_{\rm{on}}$ and the hyperfine interaction $A_{0}$. The infidelities of both intra- and inter-cluster multi-qubit gates increase due to variations in the exchange interaction induced by charge noise.}
		\label{fig:device}
	\end{figure}

    \subsection{Optimized donor number within a single cluster}\label{sec: number}
    Stochastic placement of donors can lead to randomness in the numbers of donors within each cluster. However, variation in the number of donors within a cluster is acceptable for a cluster-based qubit system. When the number of donors in the cluster unexpectedly increases, the excess nuclear spins can be excluded from the qubit set after they are initialized. A decrease in the number of donors does not disrupt the cluster array structure, as the nuclear spin qubits within the array remain interconnected via exchange interactions. Although variation in the number of donors does not affect the feasibility of the cluster-based scheme, determining the optimal range of donor numbers can ensure good operation performance of the qubits. If there are too few donors within a cluster, more inter-cluster operations will be required, reducing the overall operational precision. In contrast, if there are too many donors in a cluster, the local cost of the CNOT gate within the cluster will increase significantly. This is because conditional ESR pulses must be applied more extensively to account for all possible states, effectively shielding the influence of other nuclei. Additionally, since the HF coupling strength depends on the overlap between the corresponding nuclear and electron wavefunctions, a large number of donors can lead to similar HF strength among nuclei within the cluster. This results in frequency crowding, making individual addressing more challenging due to overlapping resonance frequencies. 
	
	To estimate the upper bound of donor numbers within a single cluster, we performed numerical simulations to model the HF strength distribution within a single cluster. Considering an ESR Rabi frequency of 0.5 MHz, a frequency difference of 5 MHz is required to ensure the crosstalk error below 1\%. We randomly sampled HF couplings from the experimentally reported range of 0.6 MHz~\cite{edlbauer2025} to 304 MHz~\cite{kranz2023use_of_exchange} until the difference between any two HF strengths was less than 10 MHz. By 1000 samplings, the feasible number of donors within a cluster ranges from 2 to 11. The average number is 4.3 (1.5). With crosstalk mitigation techniques or slower ESR driving~\cite{ZhangCH2025}, the feasible number of donors within a cluster can be further increased. 
    
    
    \subsection{Tuning up of the system}\label{sec: tuning-up} 
    The first step for system tuning up is the charge filling within each cluster. There are two key voltage working points that must be pre-calibrated: one is a region where the exchange interaction is turned off, allowing both electron spin readout~\cite{hogg2023single} and the subsequent intra-cluster calibration process; the other is where exchange interaction between neighboring clusters is turned on, which enables inter-cluster quantum operations. Moreover, given the inherent complexity of the charge stability diagram for multi-cluster devices, a complementary approach combining virtual gates~\cite{volk2019loading, mills2019shuttling} and charge transition simulations~\cite{krzywda2025qdarts} is essential for the calibration of the hyperfine (HF) interaction and exchange interaction by using the electron spin resonance (ESR) and nuclear magnetic resonance (NMR) spectrum.
    
    Based on the above, the following tuning-up procedure is performed under two stages: one with the exchange interaction off, and the other with it on. (i) Initially, the exchange interaction is turned off, allowing each cluster to be treated as a single cluster. The ESR and NMR frequencies for a single cluster are characterized as follows: first, an adiabatic ESR sweep is performed, from which the ESR frequencies and HF interactions can be extracted. These results are subsequently used to confirm the corresponding NMR frequencies. Using the measured ESR and NMR frequencies along with the protocol in Refs.~\cite{Thorvaldson2024}, the spin system can be initialized into a specific quantum state, and the exemplary initialization circuit is shown in Fig.~\ref{fig:control}(a) in the main text. This enables the calibration of NMR-based single-qubit and ESR-based intra-cluster multi-qubit gates. The entire procedure described above completes the calibration for a single cluster. Furthermore, the development of RF-multiplexing readout~\cite{ruffino2022,donnelly2025} and multi-tone driving~\cite{wuYh2025} techniques enables the simultaneous calibration of multiple clusters (with exchange interaction off), offering the potential to significantly accelerate the overall process.
    (ii) When the exchange interaction is turned on, the two adjacent clusters are coupled. In this case, each cluster pair can be treated as a ``large single cluster'' and undergoes the same calibration procedure. Since each inter-cluster operation only requires activating the exchange interaction between the corresponding adjacent clusters, it is only necessary to pre-calibrate all adjacent cluster pairs and the exchange interaction can be extracted from Chevron pattern of ESR spectrum. Thus, inter-cluster multi-qubit gates can be calibrated in the same way as above.
    
	\section{Performance evaluation and error analysis for qubit operations}
	\subsection{Analytical calculation}\label{sec:analytical}
	Analytical calculations help better understand the impact of different interactions on qubit operations, particularly the crosstalk issue when frequency detuning is insufficient. We consider cluster with $N_{\rm{L}}$ donors and a neighboring cluster with $N_{\rm{R}}$ donors, whose Hamiltonian can be expressed as
	\begin{equation}
		H_{2C} = B_{0}\left[\sum_{i}^{\rm{L,R}}\gamma_{e}S_{z}^{i}+\sum_{i=1}^{N_{\rm{L}}+N_{\rm{R}}}\gamma_{n}S_{z}^{i}\right]+\sum_{i=1}^{N_{\rm{L}}}A_{i}\mathbf{S_{L}\cdot I}^{i}+\sum_{i=N_{\rm{L}}+1}^{N_{\rm{L}}+N_{\rm{R}}}A_{i}\mathbf{S_{R}\cdot I}^{i}+J\mathbf{S_{L}\cdot S_{R}}.
	\end{equation}
	where $\gamma_{e}$ and $\gamma_{n}$ are the gyromagnetic ratio for the electron spin and the nuclear spin, $A_{i}$ is the strength of the HF interaction between electron spin and nuclear spins on the left (right) donor, and $J$ is the exchange coupling between electron spins. $\mathbf{S_{L(R)}}$ is the spin operator for the left (right) electron spin. $\mathbf{I^{i}}$ is the spin operator for the left (right) nuclear spin of nucleus $i$. The system Hamiltonian includes Zeeman terms, HF terms and exchange terms. By diagonalizing $H_{2C}$, the energies of electron spin states with different combination of nuclear spin states can be obtained:
	\begin{equation}
		\begin{aligned}
			E_{\downarrow_{\rm{L}}\downarrow_{\rm{R}},\{m_{i}\}} &= -\gamma_{e}B_{0} + \sum_{i=1}^{N_{\rm{L}}+N_{\rm{R}}}(-1)^{m_{i}}\left(\frac{\gamma_{n}B_{0}}{2} - \frac{A_{i}}{4}\right) +\frac{J}{4},\\
			E_{\downarrow_{\rm{L}}\uparrow_{\rm{R}},\{m_{i}\}} &=  \sum_{i=1}^{N_{\rm{L}}+N_{\rm{R}}}(-1)^{m_{i}}\frac{\gamma_{n}B_{0}}{2} \mp \frac{\sqrt{\Delta^{2}(A_{i})+4J^{2}}}{4}- \frac{J}{4},\\
			E_{\uparrow_{\rm{L}}\downarrow_{\rm{R}},\{m_{i}\}} &= \sum_{i=1}^{N_{\rm{L}}+N_{\rm{R}}}(-1)^{m_{i}}\frac{\gamma_{n}B_{0}}{2} \pm \frac{\sqrt{\Delta^{2}(A_{i})+4J^{2}}}{4} - \frac{J}{4},\\
			E_{\uparrow_{\rm{L}}\uparrow_{\rm{R}},\{m_{i}\}} &= \gamma_{e}B_{0}+ \sum_{i=1}^{N_{\rm{L}}+N_{\rm{R}}}(-1)^{m_{i}}\left(\frac{\gamma_{n}B_{0}}{2} + \frac{A_{i}}{4} \right)+ \frac{J}{4} ,
		\end{aligned}
	\end{equation}
	where ${n_{i}}$ represents different combinations of nuclear spins, determining the value of $m_{i}$. When the nuclear spin state is $\mid\Uparrow\rangle$ ($\mid\Downarrow\rangle$), $m_{i} = 0$ (1).  Additionally,
	\begin{equation}\label{eq:delta}
		\Delta (A_{i}) = (-1)^{m_{\rm{e,L}}}\sum_{i=1}^{N_{\rm{L}}}(-1)^{m_{i}}A_{i} +(-1)^{m_{\rm{e,R}}}\sum_{i=N_{\rm{L}}+1}^{N_{\rm{L}}+N_{\rm{R}}}(-1)^{m_{i}}A_{i}.
	\end{equation}
	Notably, $m_{\rm{e,L}}$ and $m_{\rm{e,R}}$ represent states of left and right electron spin respectively. $m_{\rm{e,L/R}} = 0$ (1) for the electron spin state $\mid\uparrow\rangle$ ($\mid\downarrow\rangle$). For the cluster-based scheme, since state of electron spin is fixed to $\mid\downarrow\rangle$ state, the NMR frequency for the $i$-th nuclear spin is
	\begin{equation}
		\hbar\omega_{i} = |\gamma_{n}B_{0} + \frac{A_{i}}{2} |. 
	\end{equation} 
	The ESR frequencies for both electron spins can be calculated:
	\begin{equation}\label{eq:hw}
		\begin{aligned}
		\hbar\omega_{\updownarrow_{\rm{L}}\downarrow_{\rm{R}},\{n_{i}\}} &= \gamma_{e}B_{0} + \sum_{i=1}^{N_{\rm{L}}+N_{\rm{R}}}(-1)^{m_{i}}\frac{A_i }{4} - \frac{J}{2} \pm \frac{\sqrt{\Delta^{2}(A_{i})+4J^{2}}}{4},\\
		\hbar\omega_{\updownarrow_{\rm{L}}\uparrow_{\rm{R}},\{n_{i}\}} &= \gamma_{e}B_{0} + \sum_{i=1}^{N_{\rm{L}}+N_{\rm{R}}}(-1)^{m_{i}}\frac{A_i }{4} + \frac{J}{2} \pm \frac{\sqrt{\Delta^{2}(A_{i})+4J^{2}}}{4},\\
		\hbar\omega_{\downarrow_{\rm{L}}\updownarrow_{\rm{R}},\{n_{i}\}} &= \gamma_{e}B_{0} + \sum_{i=1}^{N_{\rm{L}}+N_{\rm{R}}}(-1)^{m_{i}}\frac{A_i }{4} - \frac{J}{2} \mp \frac{\sqrt{\Delta^{2}(A_{i})+4J^{2}}}{4},\\
		\hbar\omega_{\uparrow_{\rm{L}}\updownarrow_{\rm{R}},\{n_{i}\}} &= \gamma_{e}B_{0} + \sum_{i=1}^{N_{\rm{L}}+N_{\rm{R}}}(-1)^{m_{i}}\frac{A_i }{4} + \frac{J}{2} \mp \frac{\sqrt{\Delta^{2}(A_{i})+4J^{2}}}{4},
		\end{aligned}
	\end{equation}
	where $\hbar\omega_{\updownarrow_{i}\downarrow_{j}}$ ($\hbar\omega_{\updownarrow_{i}\uparrow_{j}}$) represents the ESR frequency of the electron $i$ when the electron $j$ is in the spin-down (up) state. Using Eq.~\eqref{eq:hw}, the conditional ESR frequencies can be calculated for different control configurations. Then, the minimum frequency difference can be obtained to estimate the crosstalk error on ESR-based multi-qubit CZ gate under specific parameters. 

    \subsection{Error analysis in analytic estimation}\label{sec:error_anal}
	This section describes the error models considered in the work and their corresponding calculations. There are three kinds of error sources: qubit decoherence, noise-induced exchange interaction variation, and crosstalk. 
	
	Assuming the decoherence time of the nuclear spin qubit is $T_{2,n}$, the error induced by nuclear spin decoherence during the operation time $\tau$ can be approximated by 
	\begin{equation}
		e_{T_{2,n}} = 1 - \textup{e}^{-\frac{\tau}{T_{2,n}}}.
	\end{equation}
	Since electron spin qubits have long relaxation times and are only used as ancilla qubits, only the electron spin decoherence during the ESR-based CZ gate is considered. The corresponding error is represented by
	\begin{equation}
		e_{T_{2,e}} = 1 - \textup{e}^{-\frac{\tau_{\rm{ESR}}}{T_{2,e}}},
	\end{equation}
	where $\tau_{\rm{ESR}}$ is the operation time of 2$\pi$ ESR operations, and $T_{2,e}$ is the decoherence time of electron spin qubit.
	
	Charge noise leads to variations in the exchange interaction. The exchange interaction is expressed as
	\begin{equation}
		J = \frac{4t_{c}^{2}U}{U^{2}-\epsilon^{2}},
	\end{equation}
	where $t_{c}$ is the tunneling between clusters, $U$ is the on-site Coulomb interaction, and $\epsilon$ is the energy detuning of the electron between clusters. To estimate the variation due to charge noise, the first derivative of $J$ with respect to $\epsilon$ yields
	\begin{equation}
		\frac{\partial J}{\partial \epsilon} = \frac{8t_{c}^{2}U\epsilon}{(U^{2}-\epsilon^{2})^{2}}.
	\end{equation}
	Then, then variation in $J$ is
	\begin{equation}\label{eq:deltaJ}
		\delta J = \delta\epsilon \frac{\partial J}{\partial \epsilon},
	\end{equation}
	where $\delta\epsilon$ is the variation in the $\epsilon$. Then, $\delta J$ is taken into the total energy variation to obtain the modified decoherence time $T_{2,e}^{\delta J}$.
	
	   There are two types of errors associated with addressability. The first is crosstalk error, which arises from unwanted spin driving caused by frequency crowding. The second is called detuning error,caused by residual coupling that shifts the frequency and hinders proper phase accumulation on the target nuclear spin state. Due to the frequency crowding issue, crosstalk errors may arise during the driving. Assuming a frequency detuning $\Delta\hbar\omega$ and resonant driving strength $f$, the crosstalk error can be estimated as
	\begin{equation}\label{eq:cs}
		e_{ct} = \frac{f^{2}}{f^{2}+ (\Delta\hbar\omega)^{2}}.
	\end{equation}
	The frequency detuning is calculated by analytical results described in Sec.~\ref{sec:analytical}. This is overestimation since the impact of electron spin decoherence on the data qubit (nuclear spin) is indirect. This equation can also be used to estimate the driving error when there is detuning $\Delta\hbar\omega_{1}$ between the driving frequency and the target frequency:
	\begin{equation}
		e_{detuning} = 1 - \frac{f^{2}}{f^{2}+ (\Delta\hbar\omega_{1})^{2}}.
	\end{equation}
	The detuning-induced error is considered when calculating fidelities of intra-cluster multi-qubit gates with the residual exchange interaction $J_{\rm{off}}$. When considering crosstalk error in ESR-based CZ gates, the critical factor is the accurate accumulation of phase. Thus, Eq.~\eqref{eq:cs} is modified to more accurately estimate the error on the nuclear spin qubits:
	\begin{equation}
		e_{ct}^{\rm{ESR}} = 1 - \frac{1}{4}\left[1 + \sqrt{1-\frac{f^{2}}{f^{2}+ (\Delta\hbar\omega)^{2}}}\right]^{2}.
	\end{equation}
	However, it should be noted that the crosstalk error estimated by the equation remains larger than the practical situation. 
	
	Using the methods for error calculation, fidelities of quantum gates can be calculated accordingly. For example, a intra-cluster three-qubit Toffoli gate for a 3P cluster includes two NMR pulses and one ESR pulse. The fidelity can be obtained by
	\begin{equation}
		F_{\rm{Intra-Toffoli}} = (1-e_{T_{2,n}}) (1-e_{T_{2,e}}^{\delta J}) (1-e_{ct})(1-e_{detuning}).
	\end{equation}
	Similarly, the fidelity of a CNOT gate including two NMR pulses and two ESR pulses is
	\begin{equation}
		F_{\rm{Intra-CNOT}} = (1-e_{T_{2,n}}) (1-e_{T_{2,e}}^{\delta J})^{2} (1-e_{ct})^{2}(1-e_{detuning})^{2}.
	\end{equation}

    \subsection{Crosstalk errors during qubit operations}\label{sec:crosstalk_result}
    \begin{figure*}[hbt!]
		\begin{center}
			\includegraphics[width=1\columnwidth]{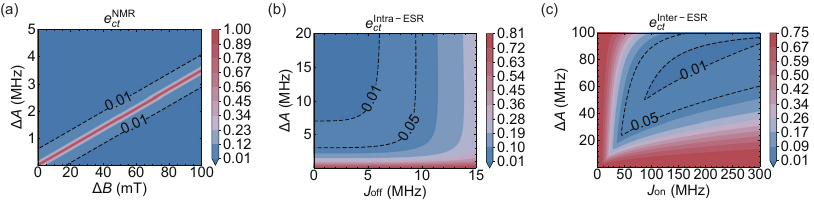}
		\end{center}
		\vspace{-0.50cm}
		\caption{Crosstalk error analysis for addressable control.
        (a) The crosstalk error rate $e_{ct}^{\rm{NMR}}$ of the NMR operation is plotted as a function of the magnetic field gradient $\Delta B$ and the minimum hyperfine difference $\Delta A$. The diagonal band indicates a resonance condition leading to high crosstalk. (b) The crosstalk error rate $e_{ct}^{\rm{Intra-ESR}}$ of the intra-cluster ESR operation is plotted as a function of residual exchange interaction $J_{\rm{off}}$ and $\Delta A$. Lower $J_{\rm{off}}$ and higher $\Delta A$ are required to suppress errors. (c) The crosstalk error rate $e_{ct}^{\rm{Inter-ESR}}$ of the inter-cluster ESR operation is plotted as a function of activated exchange interaction $J_{\rm{on}}$ and $\Delta A$.}
		\label{fig:control-Addressibility}
	\end{figure*}
    To systematically evaluate how HF interaction strengths, magnetic field gradients, and exchange interactions affect gate operations in our cluster-based architecture, we calculated crosstalk error rates during NMR, intra-cluster ESR, and inter-cluster ESR operations. The calculation method of the crosstalk error is detailed in Supplementary Materials Sec.~\ref{sec:error_anal}. The static magnetic field $B_0$ is set to be 1.35 T. For simplicity, a 2P-1P pair is taken as an example. The minimum difference between HF interactions is defined as $\Delta A$. The Rabi frequency is set to 30 kHz for NMR and 0.5 MHz for ESR, both higher than typical values to highlight the impact of crosstalk errors~\cite{edlbauer2025}. To ensure the result reflects the worst case, the crosstalk error is calculated according to the smallest frequency difference among all possible frequency pairs.

	Figure~\ref{fig:control-Addressibility}(a) shows the crosstalk errors of NMR operations as a function of the magnetic field gradient $\Delta B$ and the minimum HF difference $\Delta A$. The crosstalk error remains below 0.01 across most of the parameter space. An error rate under this threshold requires an HF strength difference as low as 0.6 MHz, without a magnetic field gradient ($\Delta B=0$). The reason for the low crosstalk error is the weak driving strength of NMR, which minimizes off-resonant effects. However, a significant increase in error rate is observed along the diagonal bright line, where the HF interaction induced NMR frequency shift cancels out that from the magnetic field induced Zeeman splitting (i.e. $\Delta\hbar\omega_{\rm{NMR}} = \frac{\Delta A}{2} - \gamma_{n}\Delta B = 0$). 
    Normally, the HF interaction is much larger than the Zeeman splitting of the nuclear spins, so that the NMR crosstalk error is not the dominant source of error.

	Figure \ref{fig:control-Addressibility}(b) shows the crosstalk error rate during intra-cluster ESR operation. Regions with low error rates ($<0.05$ and $<0.01$) are indicated by dashed black lines. To maintain the crosstalk error rate below 0.05, the HF difference $\Delta A$ should be greater than 3 MHz, and the residual exchange coupling $J_{\rm{off}}$ should be less than 9 MHz. A smaller $\Delta A$ induces frequency crowding, and worsens crosstalk error. Additionally, a stronger $J_{\rm{off}}$ couples the electron spins, resulting in the ESR frequency depending on the nuclear spin states in the adjacent cluster. The ESR frequency shift subsequently introduces off-resonant errors and prevents the accurate accumulation of the $\pi$ phase. When residual exchange coupling $J_{\rm{off}}$ is weaker than 5 MHz, the crosstalk error of the intra-cluster ESR operation remains below 0.01.

	High-fidelity inter-cluster quantum operations are essential for scalable universal control of the system. Figure \ref{fig:control-Addressibility}(c) shows the result for the inter-cluster ESR operation (corresponding to E-TCMG). Here, the degenerate frequencies caused by accidental coincidence are excluded from the analysis for simplicity and qualitative understanding. To achieve an error rate below 0.05, the activated exchange interaction strength $J_{\rm{on}}$ should be stronger than 50 MHz, and the corresponding HF difference $\Delta A$ should exceed 25 MHz. A large $\Delta A$ ensures that the ESR drive distinguishes specific nuclear spins, while a sufficiently strong $J_{\rm{on}}$ guarantees the addressability of nuclear spin states in adjacent clusters. 
    
	\subsection{Method for numerical simulation}~\label{sec: numerical}
	In this work, numerical calculations are perform by solving the Schr$\ddot{\text{o}}$dinger equation:
	\begin{equation}
		i\hbar\frac{\partial}{\partial t}|\psi(t)\rangle = H(t) |\psi(t)\rangle,
	\end{equation}
	where $H(t)$ is the time-dependent Hamiltonian for the clusters. Here, $H(t)$ is obtained by expressing the system Hamiltonian in a proper rotating frame. In this work, the rotating frame is defined with respect to the diagonalized static Hamiltonian. In such rotating frame, the resonant driving terms become time-independent:
	\begin{equation}
		d_{ij} = d_0,
	\end{equation}
	where $d_0$ is the effective strength of driving between state $i$ and state $j$. The off-resonant terms remain oscillatory:
	\begin{equation}
		d_{i'j'} = d_{0}' e^{\pm i\Delta' t},
	\end{equation}
	where $\Delta'=\hbar\omega_{ij} - \hbar\omega_{i'j'}$ is the frequency detuning. Thus, the crosstalk issue caused by frequency crowding can be effectively captured.
	
	Next, to investigate the impact of noise on qubit operations, we also incorporated noise. For simplicity, noise is incorporated through static frequency detunings $\delta_j$ of the qubits.  
    Then, the noise-induced effects are then incorporated by adding the detuning Hamiltonians of spin qubits:
	\begin{equation}
		H_{\delta,j} = \delta_{j} \frac{\sigma_{Z,j}}{2},
	\end{equation}
	where the Pauli-$Z$ matrix is corresponding to spin qubit $j$.

    \subsection{Addressability analysis of inter-cluster ESR operations and comparison with numerical method}\label{sec: addressability_supp}
Based on the analytical results above, we can discuss the direct impact of exchange interaction $J$ and the minimum HF difference $\Delta A$ on the addressability of ESR operations. ESR-based multi-qubit gates serve as primitive quantum gates in the donor cluster scheme. The addressability of ESR operations depend on the differentiation of distinct ESR frequencies. The differences between ESR frequencies can be calculated using Eq.~\eqref{eq:hw}. The frequency differences are provided by HF interaction, exchange interaction and their combined effect. We take a two-cluster pair as an example and divide the discussion into two cases. First, assume that the state of nuclear spins in the system are initialized to one product state, meaning that the minimum ESR frequency difference is mainly related to the electron spins. The noticeable frequency differences include
    \begin{equation}
    	\begin{aligned}
    		\hbar\omega_{\updownarrow_{\rm{L}}\downarrow_{\rm{R}},\{n_{x}\}}-\hbar\omega_{\downarrow_{\rm{L}}\updownarrow_{\rm{R}},\{n_{x}\}} =& \frac{1}{2}\sqrt{\Delta_{l}^{2}(A_{i})+4J^{2}},\\
    		\hbar\omega_{\updownarrow_{\rm{L}}\uparrow_{\rm{R}},\{n_{x}\}}-\hbar\omega_{\downarrow_{\rm{L}}\updownarrow_{\rm{R}},\{n_{y}\}} =& J + \frac{1}{2}\sqrt{\Delta_{m}^{2}(A_{i})+4J^{2}},\\
    		\hbar\omega_{\updownarrow_{\rm{L}}\downarrow_{\rm{R}},\{n_{x}\}}-\hbar\omega_{\uparrow_{\rm{L}}\updownarrow_{\rm{R}},\{n_{x}\}} =& -J +  \frac{1}{2}\sqrt{\Delta_{l}^{2}(A_{i})+4J^{2}},\\
    		\hbar\omega_{\updownarrow_{\rm{L}}\downarrow_{\rm{R}},\{n_{x}\}}-\hbar\omega_{\updownarrow_{\rm{L}}\uparrow_{\rm{R}},\{n_{x}\}} =& J+\frac{1}{4}\left[\sqrt{\Delta_{l}^{2}(A_{i})+4J^{2}}- \sqrt{\Delta_{m}^{2}(A_{i})+4J^{2}} \right],
    	\end{aligned}
    \end{equation}
    where $l (m) = x,y$. $\Delta_{x}$ is obtained by Eq.~\eqref{eq:delta} for the $\mid\uparrow_{\rm{L}}\downarrow_{\rm{R}},{n_{i}}\rangle$ state, while $\Delta_{y}$ corresponds to the $\mid\downarrow_{\rm{L}}\uparrow_{\rm{R}},{n_{i}}\rangle$ state. When $J\approx 0$, the two electron spins are decoupled. The frequency detuning between the two electron spin is provided by the HF interactions: $\Delta_{l/m}(A_i)/2$ or $\Delta_{l}(A_i)/4-\Delta_{m}(A_i)/4$. As $J$ increases to be larger than $\Delta_{l/m}$, the frequency differences are increased:
    \begin{equation}
    	\begin{aligned}
    		\hbar\omega_{\updownarrow_{\rm{L}}\downarrow_{\rm{R}},\{n_{x}\}}-\hbar\omega_{\downarrow_{\rm{L}}\updownarrow_{\rm{R}},\{n_{x}\}} \rightarrow& J,\\
    		\hbar\omega_{\updownarrow_{\rm{L}}\uparrow_{\rm{R}},\{n_{x}\}}-\hbar\omega_{\downarrow_{\rm{L}}\updownarrow_{\rm{R}},\{n_{y}\}} \rightarrow& 2J,\\
    		\hbar\omega_{\updownarrow_{\rm{L}}\downarrow_{\rm{R}},\{n_{x}\}}-\hbar\omega_{\updownarrow_{\rm{L}}\uparrow_{\rm{R}},\{n_{x}\}} \rightarrow& J,
    	\end{aligned}
    \end{equation}
    except for
    \begin{equation}
    	\hbar\omega_{\updownarrow_{\rm{L}}\downarrow_{\rm{R}},\{n_{x}\}}-\hbar\omega_{\uparrow_{\rm{L}}\updownarrow_{\rm{R}},\{n_{x}\}} \rightarrow 0.
    \end{equation}
    This is due to strong superposition between anti-paralleled spin states. The anti-paralleled spin states are almost degenerate, resulting in no difference between frequencies correspond to them.
    
    \begin{figure}[!htbp]
    	\begin{center}
    		\includegraphics[width=0.9\columnwidth]{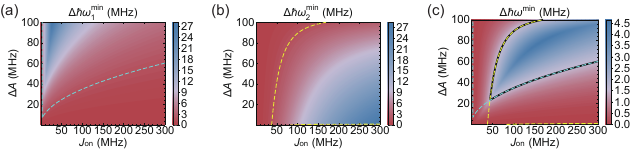}
    	\end{center}
    	\vspace{-0.50cm}
    	\caption{Minimum difference between inter-cluster ESR frequencies excluding accidental degeneracies. (a) The minimum frequency difference $\Delta\hbar\omega^{\rm{min}}_{1}$ is plotted as a function of the activated exchange coupling $J_{\rm{on}}$ and the HF difference $\Delta A$, considering differences only for different electron spin states. The cyan dashed line indicates the region where $e_{\rm{ct}}<5\%$. (b) The minimum frequency difference $\Delta\hbar\omega^{\rm{min}}_{2}$ is plotted as a function of the activated exchange coupling $J_{\rm{on}}$ and the HF difference $\Delta A$, considering differences only for different nuclear spin states. The yellow dashed line indicates the region where $e_{\rm{ct}}<5\%$. (c) The minimum frequency difference $\Delta\hbar\omega^{\rm{min}}$ is plotted as a function of the activated exchange coupling $J_{\rm{on}}$ and the HF difference $\Delta A$, considering all ESR frequencies. The black dashed line indicates the region where $e_{\rm{ct}}<5\%$.}
    	\label{fig:address_comp}
    \end{figure}
    
    Second, the degeneracy between ESR frequencies corresponding to electron spin states is ignored. Notably, the ESR-based multi-qubit gate only requires a 2$\pi$ rotation of the electron spin. The initial state of the electron spin need not be the single spin eigenstate. A 2$\pi$ rotation between the singlet-triplet states of the two electron spins on the Bloch sphere can equally achieve the target multi-qubit gate. Considering ESR driving under the condition that the spin of the right electron is down, the frequency differences include two types:
    \begin{equation}
    	\begin{aligned}
    		\hbar\omega_{\updownarrow_{\rm{L}}\downarrow_{\rm{L}},\Uparrow_{1\rm{L}},\{n_{i}/n_{1\rm{L}}\}} - \hbar\omega_{\updownarrow_{\rm{R}}\downarrow_{\rm{L}},\Downarrow_{1\rm{L}},\{n_{i}/n_{1\rm{L}}\}} &=  \frac{A_{1}}{2} + \frac{1}{4}\left[\sqrt{\Delta_{l}^{2}(A_{1\rm{L}},\{A_{i}/A_{1\rm{L}}\})+4J^{2}} - \sqrt{\Delta_{m}^{2}(-A_{1\rm{L}},\{A_{i}/A_{1\rm{L}}\})+4J^{2}} \right],\\
    		\hbar\omega_{\updownarrow_{\rm{R}}\downarrow_{\rm{R}},\Uparrow_{1\rm{R}},\{n_{i}/n_{1\rm{R}}\}} - \hbar\omega_{\updownarrow_{\rm{R}}\downarrow_{\rm{R}},\Downarrow_{1\rm{R}},\{n_{i}/n_{1\rm{R}}\}} &=  \frac{A_{1}}{2} + \frac{1}{4}\left[\sqrt{\Delta_{l}^{2}(-A_{1\rm{R}},\{A_{i}/A_{1\rm{R}}\})+4J^{2}} - \sqrt{\Delta_{m}^{2}(A_{1\rm{R}},\{A_{i}/A_{1\rm{R}}\})+4J^{2}} \right].
    	\end{aligned}
    \end{equation}
    When $J\approx 0$, the addressability of ESR frequencies is achieved via HF interaction. Therefore, the intra-cluster ESR operations remain addressable, while the inter-cluster ESR operations are impossible:
    \begin{equation}
    	\begin{aligned}
    	\hbar\omega_{\updownarrow_{\rm{L}}\downarrow_{\rm{L}},\Uparrow_{1\rm{L}},\{n_{i}/n_{1\rm{L}}\}} - \hbar\omega_{\updownarrow_{\rm{R}}\downarrow_{\rm{L}},\Downarrow_{1\rm{L}},\{n_{i}/n_{1\rm{L}}\}} &\approx A_{1\rm{L}},\\ \hbar\omega_{\updownarrow_{\rm{R}}\downarrow_{\rm{R}},\Uparrow_{1\rm{R}},\{n_{i}/n_{1\rm{R}}\}} - \hbar\omega_{\updownarrow_{\rm{R}}\downarrow_{\rm{R}},\Downarrow_{1\rm{R}},\{n_{i}/n_{1\rm{R}}\}}&\approx 0.
    	\end{aligned}
    \end{equation}
    As $J$ increases to be large than $\Delta_{l/m}$, the frequency differences become
    \begin{equation}
    	\begin{aligned}
    		\hbar\omega_{\updownarrow_{\rm{L}}\downarrow_{\rm{L}},\Uparrow_{1\rm{L}},\{n_{i}/n_{1\rm{L}}\}} - \hbar\omega_{\updownarrow_{\rm{R}}\downarrow_{\rm{L}},\Downarrow_{1\rm{L}},\{n_{i}/n_{1\rm{L}}\}} &\approx \frac{A_{1\rm{L}}}{2},\\ \hbar\omega_{\updownarrow_{\rm{R}}\downarrow_{\rm{R}},\Uparrow_{1\rm{R}},\{n_{i}/n_{1\rm{R}}\}} - \hbar\omega_{\updownarrow_{\rm{R}}\downarrow_{\rm{R}},\Downarrow_{1\rm{R}},\{n_{i}/n_{1\rm{R}}\}}&\approx A_{1\rm{R}}.
    	\end{aligned}
    \end{equation}
    Consequently, as $J$ increases, the frequency differences between intra-cluster ESR operations are decreased to $A_{1\rm{L}}/2$, while the frequency differences between inter-cluster ESR operations are increased to $A_{1\rm{R}}$.
    
    To illustrate how the HF coupling and exchange interaction determine the optimal parameter range, the minimum frequency differences under the two categories are plotted as functions of the exchange coupling $J$ and minimum HF coupling difference $\Delta A$. The corresponding results are plotted in Fig.~\ref{fig:address_comp}. Fig.~\ref{fig:address_comp}(a) and (b) compare the two kinds of the minimum ESR frequency differences $\Delta\hbar\omega_{1}^{\rm{min}}$ and $\Delta\hbar\omega_{2}^{\rm{min}}$. Then, Fig.~\ref{fig:address_comp}(c) shows the minimum ESR frequency difference $\Delta\hbar\omega^{\rm{min}}$ considering both $\Delta\hbar\omega_{1}^{\rm{min}}$ and $\Delta\hbar\omega_{2}^{\rm{min}}$. Notably, the accidental degeneracy arising from frequency crossing is neglected. Fig. \ref{fig:address_comp}(a) shows the result for ESR frequency differences corresponding to different electron spin states. As $J_{\rm{on}}$ increases from 0, the minimum of the first kind difference $\Delta\hbar\omega_{1}^{\rm{min}}$ is increased. $\Delta\hbar\omega_{1}^{\rm{min}}$ then reaches an optimal $J_{\rm{on}}^{\rm{opt}}$ point determined by $\Delta A$. As $J_{\rm{on}}$ increases beyond $J_{\rm{on}}^{\rm{opt}}$, $\Delta\hbar\omega_{1}^{\rm{min}}$ decreases. This is primarily due to that the two electron spin states gradually form spin singlet-triplet states. Consequently, the two electron spins become indistinguishable, resulting in closely spaced ESR frequencies. The result for ESR frequency differences corresponding to distinct nuclear spin states is shown in Fig.~\ref{fig:address_comp}. As $J_{\rm{on}}$ increases, the minimum of the second difference $\Delta\hbar\omega_{2}^{\rm{min}}$ is increased. In contrast, as $\Delta A$ increases, $\Delta\hbar\omega_{2}^{\rm{min}}$ is decreased. The differentiation between ESR frequencies conditional on the nuclear spin states within adjacent cluster is based on the coupling between the two electron spins, which is enhanced by exchange interaction but suppressed by the energy splittings between electron spins. In general, the minimum differences $\Delta\hbar\omega^{\rm{min}}$ in the two cases exhibit opposite dependencies on $J_{\rm{on}}$ and $\Delta A$. Consequently, the optimal region for $\Delta\hbar\omega^{\rm{min}}$, when all ESR frequencies are considered, is collectively determined by both cases, as shown in Fig.~\ref{fig:address_comp}(c).
    
    By taking the accidental degeneracy into account, we calculated the minimum ESR frequency difference considering all ESR frequencies using Eq.~\eqref{eq:hw}. To verify the validity of the analytical results, we also numerically calculated the minimum ESR frequency differences under the same conditions. The analytical and numerical results are shown in Fig.~\ref{fig:dhw_all}, showing good agreement. Compared with Fig.~\ref{fig:address_comp}(c), many dark and narrow regions with $\Delta\hbar\omega_{\rm{min}}$ close to zero appear. These arise because two ESR frequencies become accidentally degenerate under the corresponding parameters. Such degeneracies can be lifted by slightly modifying the parameters. This further demonstrates the importance of the tunability of the exchange interaction.
    
    \begin{figure}[htbp]
    	\begin{center}
    		\includegraphics[width=0.7\columnwidth]{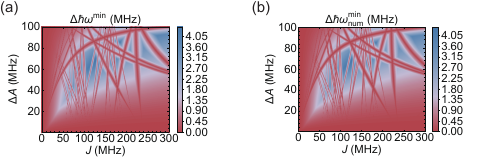}
    	\end{center}
    	\vspace{-0.50cm}
    	\caption{Minimum difference between inter-cluster ESR frequencies considering all frequencies. (a) Analytically obtained $\Delta\hbar\omega^{\rm{min}}$ is plotted as a function of $J$ and $\Delta A$. (b) Numerically obtained $\Delta\hbar\omega^{\rm{min}}_{\rm{num}}$ is plotted as a function of $J$ and $\Delta A$.}
    	\label{fig:dhw_all}
    \end{figure}
    
    \subsection{Evaluation of qubit operation performance based on numerical calculations}\label{sec:numerical-results}
	\subsubsection{Detailed analysis of addressability error by numerical method}
	\begin{figure}[htbp]
		\begin{center}
			\includegraphics[width=0.9\columnwidth]{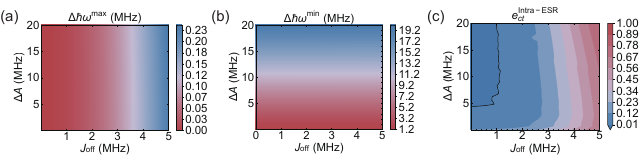}
		\end{center}
		\vspace{-0.50cm}
		\caption{Addressability analysis of the intra-cluster ESR operation obtained by numerical calculation. (a) The `maximum frequency detuning' $\Delta\hbar\omega^{\rm{max}}$ between ESR frequencies conditional on distinct states of nuclear spins in the adjacent cluster is plotted as a function of $J_{\rm{off}}$ and $\Delta A$. $\Delta\hbar\omega^{\rm{max}}$ denotes the maximum ESR frequency detuning induced by residual exchange interaction, which results in a error to acuumulate the target phase and thus reduces the gate fidelity. (b) The minimum frequency detuning $\Delta\hbar\omega^{\rm{min}}$ between ESR frequencies conditional on distinct states of nuclear spins within the cluster is plotted as a function of $J_{\rm{off}}$ and $\Delta A$. $\Delta\hbar\omega^{\rm{min}}$ represents the minimum detuning between any two ESR frequencies, which corresponds to crosstalk. (c) The infidelity $e_{ct}^{\rm{Intra-ESR}}$ of the intra-cluster ESR operation is plotted as a function of $J_{\rm{off}}$ and $\Delta A$. The black lines indicate the region with $e_{ct}^{\rm{Inter-ESR}}<1\%$.}
		\label{fig:intra_num}
	\end{figure}
    
	In this section, we numerically evaluated the qubit operation performance to provide a more accurate assessment of the impact of crosstalk on ESR operations in the cluster scheme. The method for numerical calculation is introduced in Sec.~\ref{sec: numerical}. The setup of the system is same with results shown in Sec.~\ref{sec:crosstalk_result}. A 1P-2P pair is taken as an example for simplicity. Furthermore, the minimum HF difference in a few-atom system is easier to adjust as a variable, allowing us to fully characterize its impact on crosstalk errors. First, neglecting the effect of noise, the ESR operation performance for implementing nuclear spin multi-qubit gates was calculated, showing how addressability errors affect qubit operation under different parameters. The results for the intra-cluster ESR operations are plotted in Fig.~\ref{fig:intra_num}. There are two kinds of addressability errors for the intra-cluster ESR operations: The crosstalk error is induced by crowding of ESR frequencies conditional on the distinct states of nuclear spins within the cluster. The detuning error is caused by the unwanted frequency detuning conditional on distinct states of nuclear spins in the adjacent cluster. Using Eq.~\eqref{eq:hw}, the corresponding maximum frequency detuning $\Delta\hbar\omega^{\max}$ and minimum frequency detuning $\Delta\hbar\omega^{\min}$ are calculated and shown in Fig.~\ref{fig:intra_num}(a) and (b). Then, the infidelity $e_{ct}^{\rm{Intra-ESR}}$ of the intra-cluster ESR operation is presented in Fig.~\ref{fig:intra_num}(c). In the calculation of $e_{ct}^{\rm{Intra-ESR}}$, the worst case of the two kinds of addressability errors are considered separately. The fidelities in the two cases are calculated separately, and then multiplied to obtain the total fidelity. The operation time is set to 2 $\upmu$s. As the residual exchange interaction $J_{\rm{off}}$ increases, $\Delta\hbar\omega^{\rm{max}}$ is increased, which is expected since the exchange interaction mediates the coupling between the electron spin and the nuclear spins within the adjacent cluster. Consequently, $e_{ct}^{\rm{Intra-cluster}}$ is enhanced with increased $J_{\rm{off}}$ as shown in Fig.~\ref{fig:intra_num}(c). In contrast, $\Delta\hbar\omega^{\rm{min}}$ is reduced as $\Delta A$ increases. Therefore, $e_{ct}^{\rm{Intra-ESR}}$ is suppressed with increased $\Delta A$. The dependence of $e_{ct}^{\rm{Intra-ESR}}$ with $J$ and $\Delta A$ agrees with the results obtained from the analytical calculation (see Fig.~4B in the main text).
	
	Then, the infidelity of the inter-cluster ESR operation without noise is calculated. In fact, due to the exchange-induced electron spin coupling, the effective driving strengths of the electron spins are modified under an AC magnetic field of the same intensity. To illustrate the impact of frequency differences on crosstalk errors, Fig.~\ref{fig:ct_num}(a) shows the variation of inter-cluster ESR operation infidelity with the activated exchange strength $J_{\rm{on}}$ and the HF difference $\Delta A$, under the same AC magnetic field strength (0.5 MHz) and driving time (2 $\upmu$s). In the calculations for this figure, only the superposition of the spin states corresponding to the two ESR frequencies with the smallest frequency difference is chosen as the initial state to obtain the worst case. A pulse at one of the frequencies is then applied, and the fidelity is calculated using an ideal state defined by accumulating $\pi$ phase only on the target state. Compared to Fig.~2G in the main text, the the overall trend is similar, but the infidelity is higher calculated by numerical calculation. This is partly because the analytical calculation does not directly account for phase errors, which have an upper bound of 0.5, and partly because a fixed pulse duration was considered without calibration, introducing intrinsic operational errors. Nevertheless, considering only the variation of infidelities with parameters, the two figures show good agreement, demonstrating the validity of the analytical calculations in the main text and the effect of crosstalk errors in the cluster scheme. Later on, we show that once the pulse duration is calibrated, the numerical result is better.
	
	\begin{figure}[htbp]
		\begin{center}
			\includegraphics[width=0.7\columnwidth]{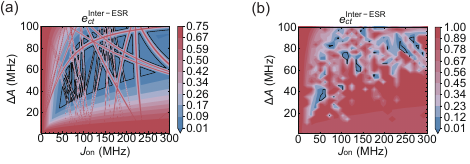}
		\end{center}
		\vspace{-0.50cm}
		\caption{Infidelity of the inter-cluster ESR operation obtained by numerical calculation. The black lines indicate the region with $e_{ct}^{\rm{Inter-ESR}}<5\%$. (a) The infidelity $e_{ct}^{\rm{Inter-ESR}}$ of the inter-cluster ESR operation is plotted as a function of $J_{\rm{on}}$ and $\Delta A$, considering fixed intensity of AC magnetic field and driving duration. (b) The infidelity $e_{ct}^{\rm{Inter-ESR}}$ of the inter-cluster ESR operation is plotted as a function of $J_{\rm{on}}$ and $\Delta A$, with calibrated operation time of 2 $\upmu$s.}
		\label{fig:ct_num}
	\end{figure}
	
	To evaluate the practical ESR operation performance, the inter-cluster ESR operation infidelities are calculated under calibrated operation times. The operation time of ESR is calibrated to 4 $\upmu$s (with distinct AC magnetic field strengths). The infidelity is plotted as a function of $J_{\rm{on}}$ and $\Delta A$ in Fig.~\ref{fig:ct_num}(b). The infidelity estimated by numerical calculation is higher than that obtained by analytical calculation. In particular, in region where $J_{\rm{on}}$ is much stronger than $\Delta A$, the infidelity is significantly higher. This is because the eigenstates of two electron spins are spin single-triplet states due to the strong exchange coupling. In this case, the ESR frequencies of the two electron spins are nearly degenerate, which enhances crosstalk errors during the ESR operation.

	\subsubsection{Noise-induced error}
	Next, we consider the effects of two types of noise: single-qubit detuning noise and exchange interaction noise due to fluctuations in th electron energy detuning. Both types of noise are modeled as static noise for simplicity. The static detuning noise for a single qubit is introduced by incorporating
	\begin{equation}
		H_{\delta,i} = \frac{\delta}{2}\sigma_{Z,i},
	\end{equation} 
	in the basis of each qubit. $\sigma_{Z,i}$ is the Pauli-Z array for qubit $i$. For exchange interaction noise, a static energy detuning $\epsilon$ is considered. Then the variation $\delta J$ is calculated using Eq.~\eqref{eq:deltaJ}. The tunneling between clusters is set to 3.6 GHz.
	
	The infidelity $1-F_{\rm{Intra-ESR}}$ for the intra-cluster ESR operation considering single-qubit detuning noise is plotted as a function of the noise strength $\delta$ in Fig.~\ref{fig:intra_noise}(a). $\delta$ is the noise strength for the electron spin qubits, while the noise strength for the nuclear spin qubits is set to $\delta/100$. We assumed that $J_{\rm{off}} \approx 0.9$ MHz and $\Delta A\approx 7$ MHz. The dependence of the infidelity is small for the typical noise amplitude ($\delta \sim 25$ kHz). With the same $J_{\rm{off}}$ and $\Delta$, $1-F_{\rm{Intra-ESR}}$ with only the exchange interaction noise is shown in Fig.~\ref{fig:intra_noise}(b). For the operational regime with weak exchange interaction, the effect of the exchange noise is not significant even when considering strong $\delta_{\epsilon}$, and as $\sigma_{\epsilon}$ increases, $1-F_{\rm{Intra-ESR}}$ decreases slightly. This is because the enhancement of the exchange interaction improves the differentiation of frequencies, thereby suppressing the dominant crosstalk error. Considering both single-qubit detuning noise ($\delta=25$ kHz) and exchange noise ($\sigma_{\epsilon}=2$ $\upmu$eV), the infidelity $e_{ct}^{\rm{Intra-ESR}}$ is shown in Fig.~\ref{fig:intra_noise}(c). Compared to the noiseless case (Fig.~\ref{fig:intra_num}(c)), the overall infidelity only increases slightly. At the specified drive strength, the crosstalk error is the dominant source of errors.
	
%
%
%
%

	\begin{figure}[htbp]
		\begin{center}
			\includegraphics[width=0.95\columnwidth]{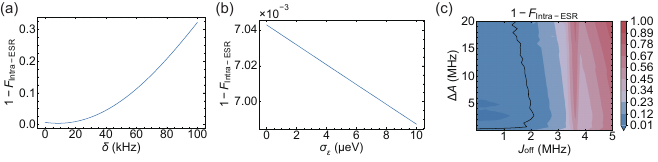}
		\end{center}
		\vspace{-0.50cm}
		\caption{Infidelity of the intra-cluster ESR operation with noise. (a) The infidelity $1-F_{\rm{Intra-ESR}}$ of the intra-cluster ESR operation is plotted as a function of static detuning strength $\delta$. (b) The infidelity $1-F_{\rm{Intra-ESR}}$ of the intra-cluster ESR operation is plotted as a function of variation on detuning $\sigma_{\epsilon}$. (c) The infidelity $e_{ct}^{\rm{Intra-ESR}}$ of the intra-cluster ESR operation is plotted as a function of $J_{\rm{on}}$ and $\Delta A$. The black lines indicate the region with $e_{ct}^{\rm{Intra-ESR}}<5\%$.}
		\label{fig:intra_noise}
	\end{figure}
	
\begin{figure}[htbp]
		\begin{center}
			\includegraphics[width=0.95\columnwidth]{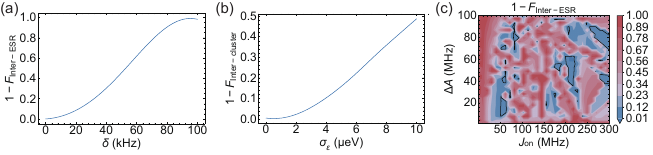}
		\end{center}
		\vspace{-0.50cm}
		\caption{Infidelity of the inter-cluster ESR operation with noise. (a) The infidelity $1-F_{\rm{Inter-ESR}}$ of the inter-cluster ESR operation is plotted as a function of static detuning strength $\delta$. (b) The infidelity $1-F_{\rm{Inter-ESR}}$ of the inter-cluster ESR operation is plotted as a function of variation on detuning $\sigma_{\epsilon}$. (c) The infidelity $e_{ct}^{\rm{Inter-ESR}}$ of the inter-cluster ESR operation is plotted as a function of $J_{\rm{on}}$ and $\Delta A$. The black lines indicate the region with $e_{ct}^{\rm{Inter-ESR}}<5\%$. When calculating the fidelity, the final states of the electron spins are neglected.}
		\label{fig:inter_noise}
	\end{figure}
	The results for the inter-cluster ESR operation are shown in Fig.~\ref{fig:inter_noise}. The infidelity with static detuning noise is presented in Fig.~\ref{fig:inter_noise}(a). We assumed that $J_{\rm{on}} = 270$ MHz and $\Delta A= 30$ MHz. Due to the slower ESR operation required for suppressing crosstalk, the infidelity $1-F_{\rm{Inter-ESR}}$ is increased more significantly compared with that for the intra-cluster ESR operation. Fig.~\ref{fig:inter_noise}(b), on the other hand, demonstrates the infidelity $1-F_{\rm{Inter-ESR}}$ when the exchange interaction noise is taken into account. The effect of the exchange noise is more significantly than that for the intra-cluster ESR operation. With a weak $\sigma_{\epsilon}$, the variation in $1-F_{\rm{Inter-ESR}}$ is negligible. As $\sigma_{\epsilon}$ further increases, the variation of infidelity $1-F_{\rm{Inter-ESR}}$ with $\sigma_{\epsilon}$ becomes more pronounced initially, and then gradually diminishes. Therefore, it is important to utilize tunable tunneling or superexchange interactions to modulate couplings between electron spins. Considering both single-qubit detuning noise ($\delta=25$ kHz) and exchange noise ($\sigma_{\epsilon}=2$ $\upmu$eV), the infidelity $e_{ct}^{\rm{Inter-ESR}}$ is shown in Fig.~\ref{fig:inter_noise}(c). Due to the more severe frequency crowding issue, both crosstalk and noise contribute to infidelity of the inter-cluster ESR operations. As discussed in the main text, the fidelity of the inter-cluster CZ gate can be further improved by utilizing singlet-triplet-based ESR operations.

\section{Multi-qubit gates for the donor-cluster system}
\subsection{Overhead of multi-qubit gates}\label{sec:overhead}
	For cluster-based nuclear spin qubits, arbitrary single-qubit gates based on NMR operations and multi-qubit CZ-type gates based on conditional ESR operations can be natively implemented. The specific implementation of the multi-qubit CZ gate is determined by the number of donors in the cluster. For a cluster consisting of $N$ donors, the number of ESR pulses required to implement a $K$-qubit CZ gate is $2^{N-K}$. The additional ESR pulses are applied to shield against the control of irrelevant nuclear spins. \added{For example, in a cluster pair with five donors, a five-qubit inter-cluster CZ gate can be natively implemented, whereas a two-qubit CZ gate between nuclear spins requires eight more ESR pulse to eliminate the dependence on the state of the non-target nuclear spins.}

    \added{Quantum error correction codes are generally based on two-qubit gates. Compared with natively realizable multi-qubit CZ gates, two-qubit CZ gates require additional ESR pulses, resulting in lower fidelity. To evaluate the performance of the two-qubit CZ gate, we make a simple estimate based on the results shown in the main text. Specifically, we take the eighth power of the corresponding fidelity in Fig.~6(c) to approximate the effect of eight ESR operations. Figure~\ref{fig:fid_CZ} shows that the infidelity of the two-qubit CZ gate remains below 1\% across a broad parameter regime. Since Fig.~6c corresponds to the worst-case scenario, Figure~\ref{fig:fid_CZ} provides a lower bound on the fidelity.}
    
    \begin{figure}[htbp]
		\begin{center}
			\includegraphics[width=0.4\columnwidth]{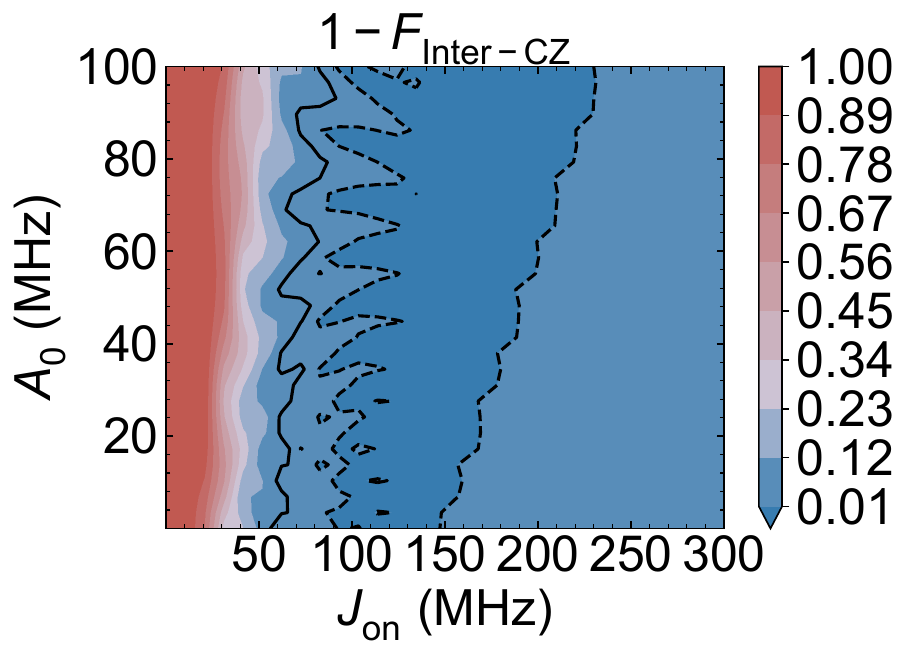}
		\end{center}
		\vspace{-0.50cm}
		\caption{The infidelity $1-F_{\mathrm{Inter-CZ}}$ of the NA-TCMG-based two-qubit CZ gate versus the activated exchange interaction $J_{on}$ and hyperfine interaction $A_{0}$. The black dashed (solid) lines indicate the region with $1-F_{\mathrm{Inter-CZ}}<1\%$ ($<5\%$).}
		\label{fig:fid_CZ}
	\end{figure}
    
    By combining two single-qubit $H$ gates and one multi-qubit CZ-type gate, a Toffoli/CNOT-type gate between nuclear spins can be implemented. To implement an $L$-qubit inter-cluster CNOT gate, when the number of donors within a cluster pair is $M$, the number of ESR pulses required is $2^{M-L}$. For example, in a 2P-3P cluster pair, a five-qubit inter-cluster Toffoli gate is implemented by two NMR pulses ($H$ gates) and one ESR pulse. However, an inter-cluster CNOT gate requires two NMR pulses and eight ESR pulses. The difference in the number of pulses required for CNOT and Toffoli gates is a notable characteristic of cluster-based nuclear spin qubits. This necessitates a tailored circuit design for algorithm demonstrations and QEC schemes. 
    The various multi-qubit gates described above ensure all-to-all connectivity among nuclear spin qubits both within a single cluster and across neighboring clusters. This local enhanced connectivity is a distinctive feature of the cluster-based qubit scheme and offers advantages in QEC, which is discussed in Sec.~\ref{sec:discussion}.

\subsection{NMR/ESR-assisted inter-cluster CZ gate}\label{sec:assist}
        \begin{figure}[htbp]
		\begin{center}
			\includegraphics[width=0.8\columnwidth]{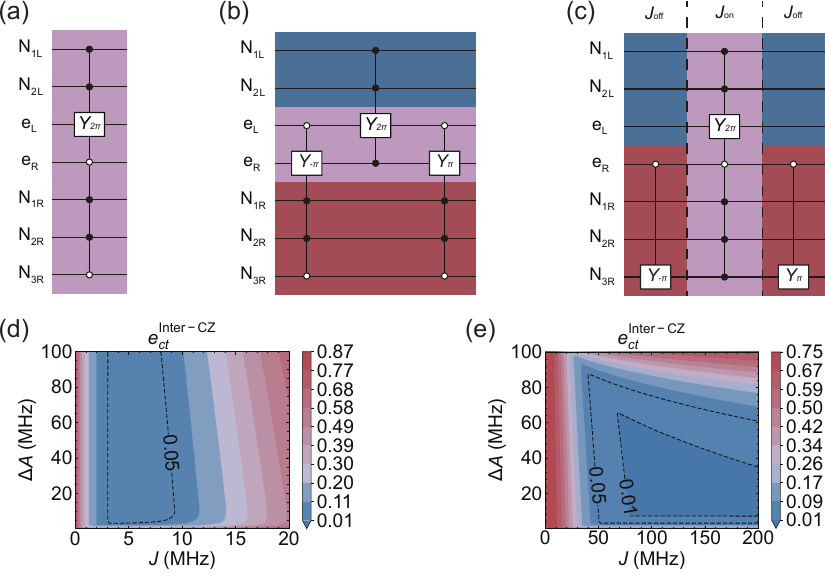}
		\end{center}
		\vspace{-0.50cm}
		\caption{EA-TCMG-based and NA-TCMG-based inter-cluster CZ gate. (a) Circuit for the single-step implementation of the target CZ gate. The geometric phase is accumulated on $\mid\Downarrow_{\rm{1}}\Downarrow_{\rm{2}}\Downarrow_{\rm{3}}\Downarrow_{\rm{4}}\Uparrow_{\rm{5}}\rangle$. (b) Circuit for the EA-TCMG of the target CZ gate. (c) Circuit for the NA-TCMG of the target CZ gate. (d) Crosstalk error $e_{ct}^{\rm{Inter-CZ}}$ for the EA-TCMG-based CZ gate is plotted as a function exchange interaction $J$ and the minimum difference between HF coupling strengths $\Delta A$. (e) Crosstalk error $e_{ct}^{\rm{Inter-CZ}}$ for the NA-TCMG-based CZ gate is plotted as a function exchange interaction $J$ and the minimum difference between HF coupling strengths $\Delta A$.}
		\label{fig:assist}
	\end{figure}
	 Although our proposed scheme can mitigate the frequency crowding issue, it may still arise for large-scale qubit expansion, leading to crosstalk. For intra-cluster operations, crosstalk errors can be suppressed by control-based methods like weaker ESR driving. However, for the inter-cluster multi-qubit CZ gates, the frequency crowding issue becomes more severe, making it difficult to suppress crosstalk. A practical alternative is to implement inter-cluster CZ gates indirectly. There are two alternative approaches for indirect nuclear-nuclear multi-qubit gates. A 2P-3P cluster pair serves as an example to illustrate these two approaches. The core of both schemes is to use additional operations to ensure that ESR operations are executed at low-crosstalk frequencies.
	 
	 The first approach is assisted by auxiliary ESR operations~\cite{edlbauer2025} and is referred to as EA-TCMG (ESR-assisted two-cluster multi-qubit gate). Notably, the EA-TCMG-based implementation requires a moderate exchange interaction to ensure that the ESR frequency is conditioned on the state of the adjacent electron spin, while remaining unaffected by nuclear spins in neighboring clusters. In the EA-TCMG-based CZ gate scheme, ESR operation are no longer required to be conditioned on adjacent nuclear spin states, thereby reducing the addressing difficulty. For example, consider a target operation that is a multi-qubit CZ-type gate on $N_{1}$ conditioned on the combined state of the other nuclear spins being $\mid\Downarrow_{2}\Downarrow_{3}\Downarrow_{4}\Uparrow_{5}\rangle$, this may be difficult to implement directly due to frequency crowding. The circuit for the direct implementation of the operation is shown in Fig.~\ref{fig:assist}(a). This issue can be addressed by the EA-TCMG-based implementation shown in Fig.~\ref{fig:assist}(b): The first step is applying a conditional $\pi$-pulse on the right electron conditioned on the nuclear spin state $\mid\Downarrow_{3}\Downarrow_{4}\Uparrow_{5}\rangle$ within the right cluster, thereby transferring the condition on nuclear spins to the electron spin. Subsequently, a $2\pi$ rotation on the left electron spin is applied conditional on the state $\mid\Downarrow_{1}\Downarrow_{2}\uparrow_{\rm{R}}\rangle$, effectively achieving the target CZ gate. Finally, the entire operation is completed by rotating the right electron spin back to the spin-down state.
	 
	 The second approach is assisted by NMR operations with a tunable $J$ and is referred to as NA-TCMG (NMR-assisted TCMG). Similar with the direct implementation, a strong exchange interaction is required to allow the ESR operation conditioned on adjacent nuclear spin states. Typically, the ESR frequencies corresponding to the all-up and all-down nuclear spin states are addressed relatively well. Other nuclear spin states can be converted to the all-up or all-down state via NMR operations to perform low-crosstalk ESR operation, and then restored to their original states using NMR. Therefore, we can combine NMR $\pi$-rotations with a multi-qubit CZ gate conditioned on the all-up (all-down) state to implement the target multi-qubit gate. The corresponding circuit is shown in Fig.~\ref{fig:assist}(c). Again, we use the same target CZ gate to illustrate this approach. The $\mid\Downarrow_{1}\Downarrow_{2}\Downarrow_{3}\Downarrow_{4}\Uparrow_{5}\rangle$ state can be transformed into the all-down state ($\mid\Downarrow_{1}\Downarrow_{2}\Downarrow_{3}\Downarrow_{4}\Downarrow_{5}\rangle$) with fewer NMR operations than into the all-up state. Thus, an NMR $\pi$-rotation is applied to the qubit $\rm{N}_{5}$. Then, an ESR 2$\pi$-rotation conditioned on the all-down nuclear spin states is applied. Finally, an inverse NMR $\pi$-rotation is applied to the qubit $\rm{N}_{5}$, effectively achieving the target nuclear-nuclear CZ gate.
	 
	 Both approaches can effectively mitigate crosstalk errors caused by frequency crowding. We take a cluster pair consisting of $b$ nuclei as an example, and the smaller number of nuclei within clusters is $a$  When implementing an $m$CZ gate ($m$ is the number of qubits involved), the EA-TCMG requires $2^{a-1}$ additional ESR operations compared to the direct implementation. In contrast, the NA-TCMG requires $2^{b-2}$ additional NMR operations compared to direct implementation. The EA-TCMG typically requires fewer additional operations compared with the NA-TCMG. Considering the faster ESR operation speed, this approach results in a short circuit execution time. However, when the exchange interaction $J$ is turned on to implement the CNOT gate between electron spins, the driving strength must be sufficiently strong to avoid being affected by the nuclear spin states in the adjacent cluster. Such strong ESR driving is prone to inducing crosstalk errors, especially in large-scale qubit systems. The NA-TCMG offers the advantage of higher operational precision and is less likely to introduce additional crosstalk. However, this approach requires longer operations times, particularly when the target nuclear spin state is far from both the all-up and all-down nuclear spin states. In summary, both schemes present distinct trade-offs. Therefore, the choice between the two approaches should be determined by the specific experiment requirements.
    
	 To compare the performance of inter-cluster CZ gates based on different implementation methods, we calculated the crosstalk errors $e_{ct}^{\rm{Inter-CZ}}$ for both the EA-TCMG-based and NA-TCMG-based inter-cluster ESR operation. The parameters are the same as those used in Sec.~\ref{sec:crosstalk_result}. The crosstalk error $e_{ct}^{\rm{Inter-CZ}}$ when applying the EA-TCMG-based CZ gate is plotted as a function of the exchange interaction $J$ and the minimum HF difference $\Delta A$ in Fig.~\ref{fig:assist}(d). In this scheme, the desirable strength of the exchange interaction is approximately from 3 to 10 MHz. This is because nuclear spin qubits within adjacent clusters are no longer required to serve as control qubits. Instead, the dependence on them of ESR operations should be suppressed. The exchange interaction must be neither too strong nor too weak. The result for the NA-TCMG is shown in Fig.~\ref{fig:assist}(e). An exchange interaction stronger than 50 MHz is sufficient. For the NA-TCMG, only the ESR on a single nuclear spin state is required. Therefore, the frequency difference increases with the enhancement of both the exchange interaction and the HF difference, thus reducing crosstalk errors. In both schemes, the requirement for the HF difference is stringent. These two schemes serve as optional complements for the control protocol for the cluster-based qubit scheme, and can be selectively employed based on the specific ESR frequency distribution within the device and the particular circuit design.

\end{widetext}

	
\end{document}